\documentclass[twocolumn, twocolappendix]{aastex631}

\begin{document}

\title{Whither or wither the Sulfur Anomaly in Planetary Nebulae?}

\author{Shuyu Tan}
\affiliation{The Laboratory for Space Research, Faculty of Science, The University of Hong Kong, Cyberport 4, Hong Kong}

\author{Quentin A. Parker}
\affiliation{The Laboratory for Space Research, Faculty of Science, The University of Hong Kong, Cyberport 4, Hong Kong}


\begin{abstract}

We present a thorough investigation of the long standing sulfur anomaly enigma. Our analysis uses 
chemical abundances from the most extensive 
dataset available for 126 planetary nebulae (PNe) with improved accuracy and reduced uncertainties 
from a $10 \times 10$ degree Galactic bulge region. By using argon 
as a superior PNe metallicity indicator, the anomaly is significantly reduced and better constrained. 
For the first time in PNe we show sulfur $\alpha$-element lock-step
with both oxygen and argon. We dispel hypotheses that the anomaly originates from 
underestimation of higher sulfur ionization stages. Using a machine learning approach, 
we show that earlier ionization correction factor (ICFs)
schemes contributed significantly to the anomaly. We find a correlation between 
the sulfur anomaly and the age/mass of PNe progenitors, with the anomaly either 
absent or significantly reduced in PNe with young progenitors. Despite inherent 
challenges and uncertainties, we link this to PNe dust chemistry, noting those 
with carbon-dust chemistry show a more pronounced anomaly. By integrating these findings, 
we provide a plausible explanation for the residual, reduced sulfur anomaly and propose its potential as an indicator of relative galaxy age compositions based on PNe.

\end{abstract}


\section{Introduction} \label{sec:intro}
Sulfur, as an $\alpha$ element, should be produced in lockstep with others like 
oxygen, neon, argon, and chlorine in more massive stars, so its cosmic abundance should also be proportional. 
Strong correlations between sulfur and oxygen abundances are seen in H~{\sc ii} regions and blue compact galaxies. 
However, historically, Planetary Nebula (PNe hereafter) sulfur abundances, which arise from low-to-intermediate mass 
progenitors, have consistently been lower, giving rise to the so-called ``sulfur anomaly'' first identified in PNe by \citet{henry2004sulfur}. 

Extensive efforts to explain the PNe sulfur anomaly through observational and theoretical approaches has so far proved 
elusive. One suggestion is past failure to properly account for the abundances of unseen ionization stages above S$^{2+}$. 
Another factor is the difference in evolutionary stages in H~{\sc ii} regions, blue compact galaxies and PNe. H~{\sc ii} 
regions are the current sites of chemical enrichment, from which the next generation of stars emerge. In contrast, PNe 
form a highly diverse and complex family, with progenitor masses spanning from $\sim$8~$M_{\odot}$ to 1 $M_{\odot}$  
and so exhibiting a wide range of ages, from tens of millions of years to over 10 Gigayears.

\citet{2013MNRAS.431.2861S} looked at whether the s-process is responsible for 
destroying sulfur during late stage stellar 
evolution but their study did not support this. Another idea is 
dust depletion of sulfur \citep{henry2012curious,2014A&A...567A..12G}. 
\citet{delgado2015oxygen} suggested oxygen and neon may be enriched 
during PN progenitor evolution which could give more scatter, 
depending on progenitor mass (and so effectively age). As a result they 
consider argon and chlorine more reliable metalicity indicators. 
Given difficulties in measuring weak chlorine lines, argon is by far 
the easier to work with and is adopted here.

{\bf \subsection{The context of the sulphur anomaly}}
Our Galaxy is an ideal testing ground for refining chemical evolutionary models. 
Galactic H~{\sc ii} regions represent zero-age systems while PNe represent older to considerably older populations 
of intermediate- to low-mass stars for study. For PNe the observed abundances 
can deviate to sub- and super-solar depending on progenitor age.  
PNe allow accurate determination of abundances among 
various chemical tracers through their emission lines, including precise 
abundance measurements of alpha elements. 
These elements, created by massive stars, remain unaltered by their progenitors and 
reflect molecular cloud composition when the stars were 
formed. If $\alpha$ elements like oxygen, neon, sulfur, and argon are 
produced in lockstep in massive stars, their abundances should exhibit equivalent 
correlations. H~{\sc ii} region and PNe observations have revealed strong correlations 
between oxygen and neon. They are considered indicators of their formation 
environments. Similar lockstep behaviour has been noted for argon. However, 
previous sulfur abundance studies in Galactic disk PNe, e.g. \citep{henry2004sulfur} 
and \citep{henry2012curious}, and Galactic Bulge PNe by \citep{cavichia2010planetary}, 
report data points that considerably deviate from the narrow, linear tracks 
for H~{\sc ii} regions. Instead they exhibit a broad spread in sulfur 
abundances that generally fall lower than for H~{\sc ii} regions. This 
discrepancy is the sulphur anomaly. It has been discussed in detail 
by e.g. \citet{henry2004sulfur} and \citet{henry2012curious} and
has been based on O/H values as the metalicity indicator. 

We examined steps from previous studies that derived
PNe sulfur abundances from their spectra. These directly determine the 
ionic abundances of S$^{+}$ and S$^{2+}$, when observable, and then multiplied by 
an ``ionisation correction factor" (ICF) e.g. 
\citep{1985ApJ...291..247M, 2014MNRAS.440..536D}, calculated to account for 
unobserved ionisation stages. Possible reasons given for the sulfur 
anomaly include underestimation of higher ionisation stages when using these 
ICFs (see later). \citet{henry2012curious} tested the sulfur 
abundance deficit across a range of nebular properties but found no strong 
correlations. Recently, by studying the born-again nebula V4334~Sgr and 
using ESO VLT spectroscopic observations from 2006 to 2022, 
\citet{reichel2022recombination} confirmed an [S~{\sc iii}] level, contrary 
to expectation of a decrease, because the S$^{2+}$ stage, if it dominates, will 
rapidly recombine to S$^{+}$. A possible explanation is recharge from underestimated 
fractions of higher ionisation stages. They propose this could explain 
the anomaly but this object is itself anomalous and only a single data point.

\section{ESO VLT FORS2 Spectroscopy} 
Our new catalogue of abundances for Galactic 
bulge PNe from quality VLT/FORS2 spectra, \citet{tan2023catalogue} 
(Paper~III) provides the most comprehensive and reliable PNe abundances available 
to re-examine the sulfur anomaly.  We found the first clear example of a lock-step relation 
for argon and neon with oxygen unlike previous studies, 
reflecting a reduction in our data's error dispersion. We re-examined this lockstep behaviour 
and the relationship between sulfur and oxygen abundances using our high S/N data from the same 
telescope and instrument and for a well-defined PNe sample.  We use these new data to investigate the  
sulfur anomaly. Our PNe selection was based on likely Galactic bulge membership, 
as detailed in Paper~I \citep{tan2023morphologies}. 

To derive abundances, we used the \citet{fitzpatrick1999correcting} extinction 
law to account for interstellar reddening. We used the ICF scheme in \citet{delgado2014ionization}, 
hereafter DMS14, to correct 
for unobserved sulfur ionisation stages. This is a key difference to previous work, e.g. 
\citet{kingsburgh1994elemental} and \citet{kwitter2001sulfur}, hereafter KW01, as used by HSK12. 
We derive S/H and O/H values for 124 objects on the standard 12~+~$\log$(X/H) scale, 
yielding high-quality results and $<0.2$~dex uncertainties.

\section{The sulfur anomaly and a comparison with literature studies}
\label{sec:res}
\begin{figure*}
    \centering
      \includegraphics[height=0.2785\textwidth]{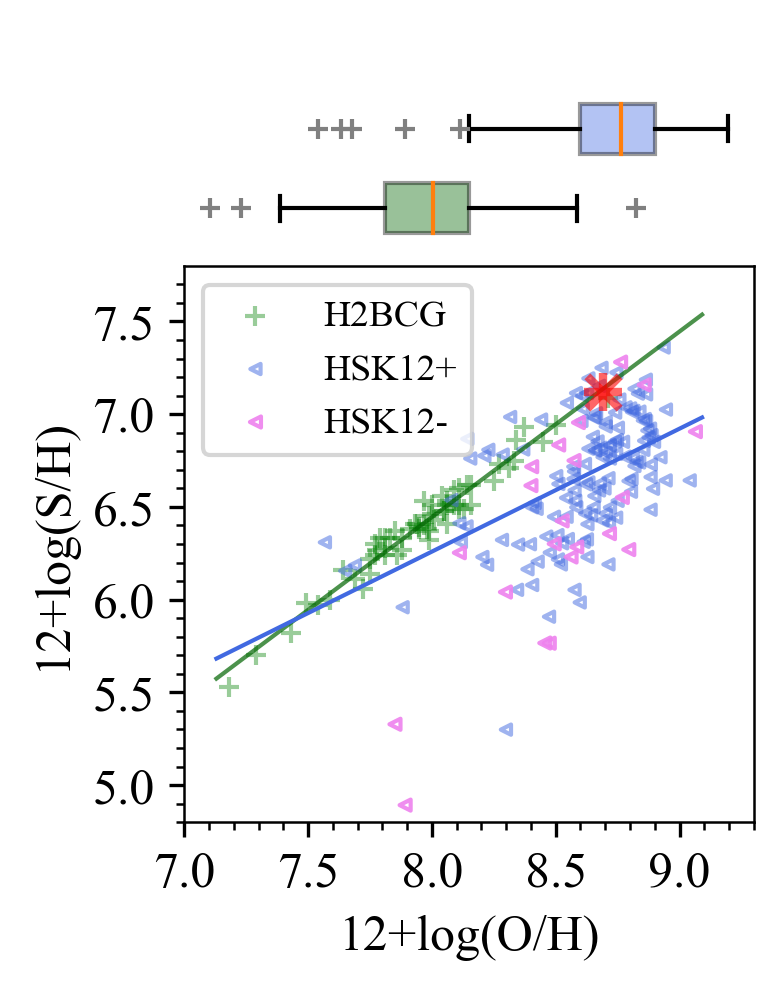}
    \includegraphics[height=0.2785\textwidth]{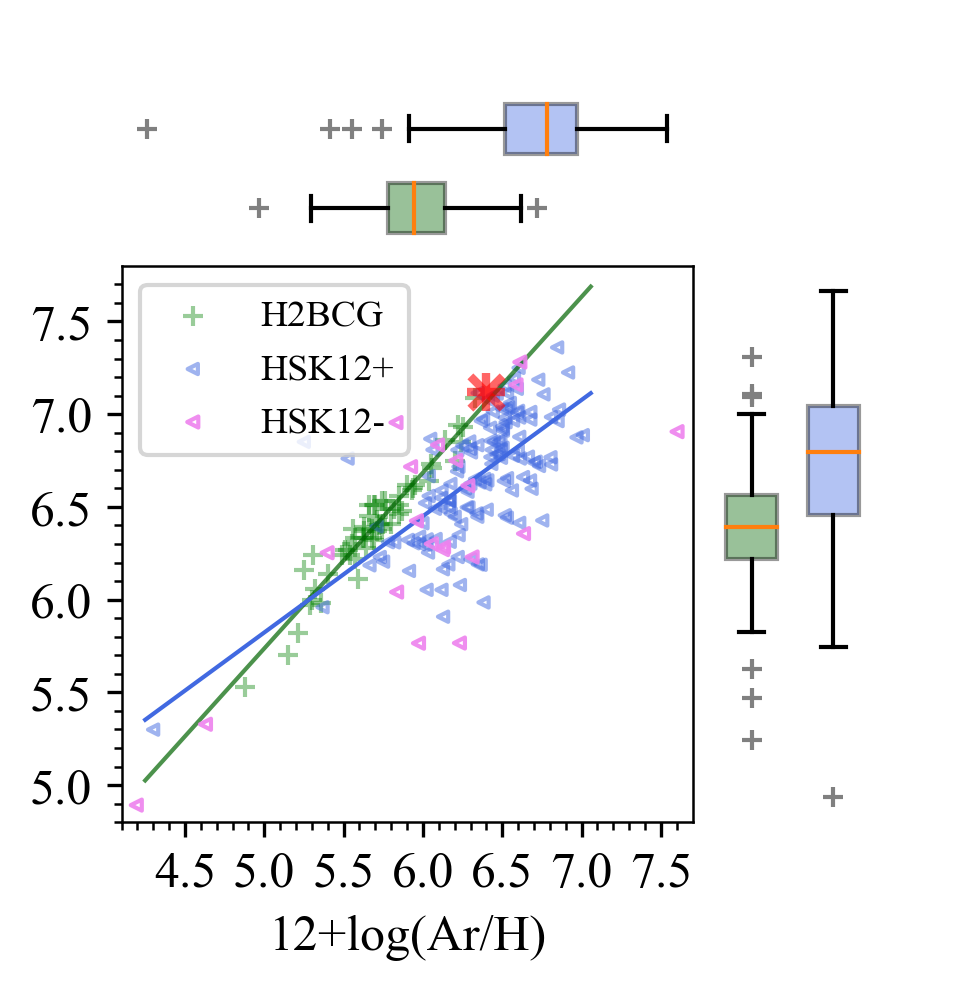}
    \includegraphics[height=0.2785\textwidth]{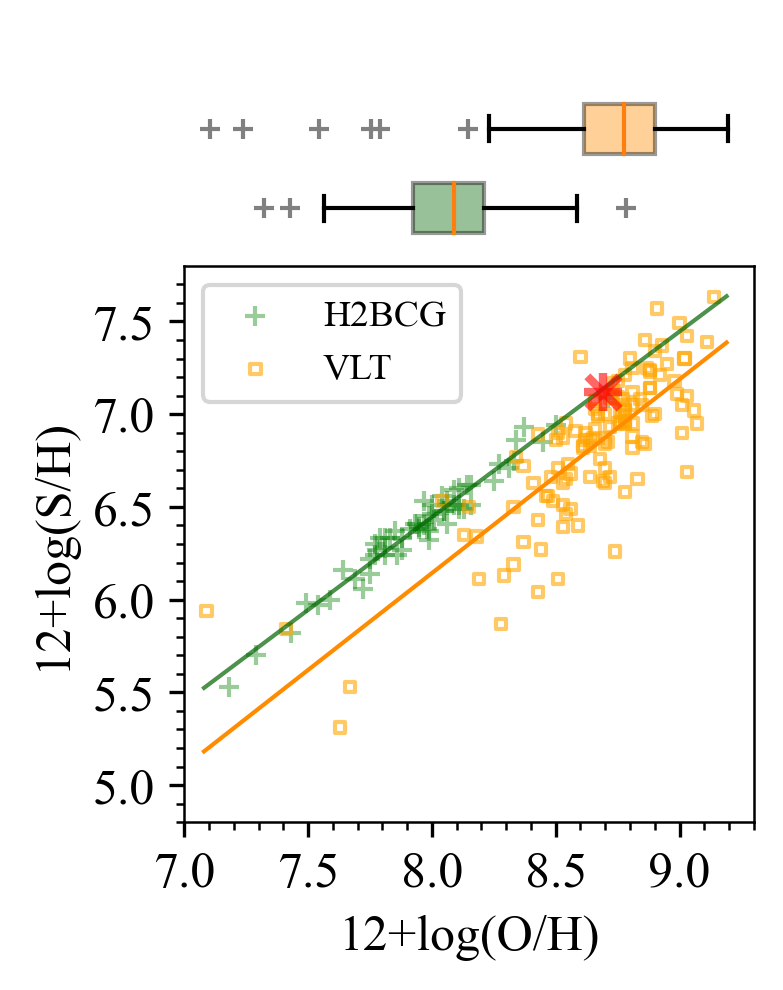}
    \includegraphics[height=0.2785\textwidth]{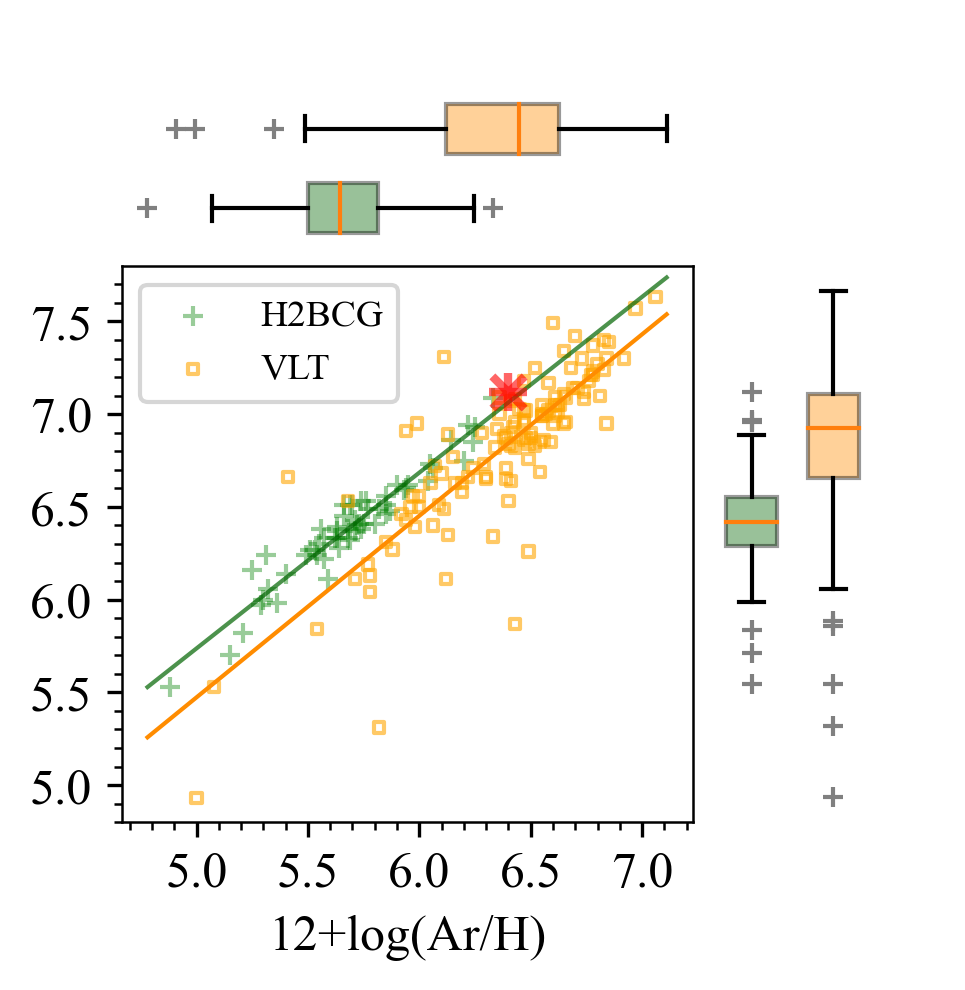}
     \caption{Comparison of S/H versus O/H and Ar/H among PNe samples 
     c.f. H2BCG data (green crosses). The two left panels show 
     Galactic disk PNe from HKS12, with blue and pink triangles triangles for high and lower quality measurements  with uncertainties $<0.2$~dex  and $>0.2$~dex. The right two panels give our Galactic bulge PNe results from 
     Paper~III as orange squares. In all panels, best-fit lines are the same color as the data. 
     Only HSK12 data points with uncertainties $<0.2$~dex (HSK12+) were fitted. 
     Our VLT results for O/H and Ar/H display best-fit lines parallel to the H2BCG trends with clear lock-step behaviour. The red asterisk is solar abundance. Color-coded boxplots at the 
    top and right of each panel depict distributions of O/H, 
    Ar/H and S/H. Red lines indicate median values with boxes extending from 
    the 25th to 75th percentiles and whiskers from 1\% to 99\% with outliers as crosses.
}
    \label{fig:sh_oh}
\end{figure*}

Fig.~\ref{fig:sh_oh} shows the relationship between S/H and O/H for 162 Galactic 
disk PNe from \citet{henry2012curious}, hereafter HSK12, and 
the 124 PNe in our ESO VLT bulge sample. HSK12 incorporates data from \citet{henry2004sulfur}, 
\citet{milingo2010alpha}, and \citet{henry2010abundances}, which adopt the \citet{kwitter2001sulfur} ICF scheme.

As proposed by \citet{delgado2015oxygen}, AGB star evolution may modify nebular oxygen 
abundances compared to their progenitors. Their analysis suggested using Ar/H or Cl/H 
as a more reliable metalicity indicator. Given the large uncertainties associated 
with Cl/H derivation, we adopt Ar/H. In Fig.~\ref{fig:sh_oh} we plot S/H against Ar/H. 
We  include data for H~{\sc ii} regions in the nearby spiral galaxy M101 from
\citet{kennicutt2003composition} and metal-poor blue compact galaxies from \citet{izotov1999heavy} 
(collectively hereafter H2BCG), where the $\alpha$-element lockstep behaviour is 
demonstrated. We exclude data from H~{\sc ii} regions with an uncertainty in $\log$(S/O)~$>0.1$~dex, as they 
exhibit a noticeable deviation from the main H2BCG data. Table~\ref{tab:fit_params} presents best-fit 
parameters for S/H versus O/H and Ar/H of each sample from linear least-squares 
regression using \texttt{scipy.stats.linregress}, together with Pearson's correlation coefficients ($r$).


The left-most panel of Fig.~\ref{fig:sh_oh} presents the non-lockstep behaviour of
S/H versus O/H for Galactic disk PNe from HSK12 (as blue and purple data points) as two trends: 
a sparsely populated (13\%) upper trend that aligns with the H2BCG data (green points), and a well-populated 
(87\%) locus below the H2BCG trend encapsulating the anomaly. The slope of the best fit line (in blue) 
significantly deviates from unity with a formal $R^{2}=0.26$, providing the main anomaly locus. 

We calculate the sulfur deficit, $d$(S), as the discrepancy between the 
predicted $12+\log$(S/H) values (from the best-fit line for $12+\log$(S/H) versus $12+\log$(O/H) of H2BCG) 
and the actual observed PNe values, following HSK12. Using only results in HSK12 with reported measurement 
uncertainty $<0.2$~dex (denoted as HSK12+), the median $d$(S) is $-0.40^{+0.31}_{-0.29}$~dex. The sulfur 
deficit seen in the S/H versus O/H plot is at least partially due to the ICF scheme used as we show in the 
next section. 

The middle-left plot of Fig.~\ref{fig:sh_oh} is the equivalent S/H versus Ar/H plot. 
Here, there is only a single, broad locus of PNe data points falling below 
but closer to the H2BCG line. This shows that using the O/H metallicity indicator 
is a key factor in the sulfur anomaly in HSK12. The data scatter in the left panels 
may reflect uncertainties in determining reliable PNe chemical abundances using 
low-resolution spectroscopic observations on 2-m telescopes. Our Paper~III shows 
that literature results based on low-resolution spectra from 2-m telescopes generally show a deviation 
$>0.1$~dex from those from the VLT observations for the $\alpha$ elements.


In the right panels of Fig.~\ref{fig:sh_oh}, we show Galactic bulge PNe data from Paper~III that displays a 
broader metalicity range, with many objects having super-solar O/H and Ar/H. Compared to HSK12, our high-quality VLT/FORS2 abundances show less scatter with most points on a narrower track below the 
H2BCG trend. The bulge PNe show clear lockstep behaviour for S/H versus O/H with a strong correlation 
coefficient of $r = 0.84$. The best-fit line has a near unity slope. This is the first time lockstep 
behaviour between S/H and O/H has been shown for any PNe sample. The S deficit is a smaller systematic 
offset of $\sim0.2$~dex, with a median $d$(S) of 
$0.24_{-0.18}^{+0.27}$~dex. The scatter at the sub-solar O/H remains significant. The sulfur lockstep 
behaviour is further tightened in S/H versus Ar/H, as in the right-most panel of Fig.~\ref{fig:sh_oh}. These 
are our major new findings.

We reanalysed the HSK12 chemical abundances using their de-reddened line intensities following  methods described in 
Paper~III\footnote{ Due to modest blue resolution of the HSK12 spectra ($\sim 1.4$~\AA), the [Ar~{\sc IV}]~$\lambda$4711 
and He~{\sc I}~$\lambda$4713 lines, and the [O~{\sc II}]~$\lambda\lambda$3726+28 doublet, cannot be resolved. Hence, no 
electron densities could be determined for [Ar~{\sc IV}] or [O~{\sc II}]. We also excluded [Ar~{\sc IV}]~$\lambda$4711 
from calculating the Ar$^{3+}$ abundances.}. Our recalculated values show median differences compared to HSK12 for S/H and 
O/H of 0.1 and $-0.07$ dex, respectively, while agreeing within the measurement uncertainty for Ar/H. These discrepancies 
could be attributed to the differences in atomic datasets and ICF schemes used. As a result of our recalculations, the 
median $d$(S) reported in HSK12 and shown in Figure~\ref{fig:sh_oh_ML} would be halved to $0.19_{-0.27}^{+0.24}$ dex.

\begin{table*}
    \centering
\begin{tabular}{lcccccccccc}
\hline\hline
\multirow{2}{*}{Sample} & \multirow{2}{*}{X/H vs. S/H} & \multicolumn{4}{c}{S/H from ICF scheme} && 
\multicolumn{4}{c}{S/H from ML} \\ \cline{3-6} \cline{8-11} & 
&slope&\phantom{$-$}intercept&$R^{2}$&$r$&&slope&\phantom{$-$}intercept&$R^{2}$&$r$  \\ \hline
VLT [124]    & O &  1.05$\pm$0.06    &
               $-$2.22$\pm$0.53  & 
               0.71 &  0.84 &&  0.99$\pm$0.06 &  $-$1.70$\pm$0.51\phantom{$-$}  & 0.70 & 0.84 \\    
        & Ar&  0.98$\pm$0.06 &  
               \phantom{$-$}0.58$\pm$0.38    &  
               0.69  &   0.83  &&   0.95$\pm$0.06      &  0.75$\pm$0.38  & 0.70   & 0.83 \\
HSK12+ [140] & O &  0.66$\pm$0.10  &
              \phantom{$-$}0.96$\pm$0.82 &
               0.26 & 0.51  &&   0.78$\pm$0.11 &  $-$0.02$\pm$0.93\phantom{$-$}  & 0.29 & 0.54 \\
        & Ar&  0.63$\pm$0.06  &   
              \phantom{$-$}2.69$\pm$0.38   &   
              0.44  &  0.67   &&  0.70$\pm$0.07     &   2.31$\pm$0.44  & 0.44  & 0.67 \\
H2BCG [61]  & O &   1.00$\pm$0.03   & $-$1.56$\pm$0.21& 0.96 & 0.98 &  &  -  & -  & - & - \\
        & Ar  &   0.95$\pm$0.04  &  \phantom{$-$}1.00$\pm$0.21 &    0.92  & 0.96 && -  & -  & -   & - \\ \hline
\end{tabular}
\caption{Slope, intercept , corresponding errors, and the 
coefficient of determination ($R^{2}$) from the least-squares regression 
for the $12+\log$(S/H) versus both $12+\log$(O/H) and $12+\log$(Ar/H) data, along with correlation coefficients ($r$) 
for Galactic disk PNe from HSK12+ and bulge PNe from Paper~III, as shown in Fig.~\ref{fig:sh_oh}. 
The best-fit parameters derived with S/H from the literature ICF schemes are presented in cols.~3-6 
while those obtained with S/H from machine learning-derived icf(S) are presented in cols.~7-10. 
The final two rows presents results for H2BCG.}
    \label{tab:fit_params}
\end{table*}


\section{Understanding the sulfur anomaly}
\label{sec:confirm_SA}
We examine the sulfur anomaly in Galactic disk and bulge PNe samples. 
Despite a notable reduction in the S/H deficit found in our high-quality bulge PNe data, the 
sulfur anomaly persists for some PNe. We focus on potential artificial causes of this anomaly, 
including underestimation of sulfur higher ionization stages beyond S$^{2+}$ 
in photoionization models. We implement a machine learning approach to derive an ad hoc 
sulfur ICF for each object to examine the ICF scheme proposed in the literature. We compare our optical 
observations with infrared literature results where higher ionisation stages could be directly observed, 
to examine possible issues associated with use of optical observations.

\subsection{Validation of ICF scheme via machine learning}
The KW01 and DMS14 ICF literature formulae were obtained through analytical 
fitting to photoionization models. The sulfur anomaly has been 
attributed to failure to properly account for high ionisation stages through use of an incorrect 
ICF for sulfur in HSK12, something  our re-calculations of their data confirm. \cite{2022MNRAS.511....1S} 
first introduced machine learning (ML) techniques to 
determine ICFs tailored to individual PNe. Here, we apply ML to a large sample. To 
minimise uncertainties with the ICF formulae, we adopt a random forest (RF) regressor 
\citep{breiman2001random} (a learning method based on decision tree regression), 
to predict ICF values (S/(S$^{+}$+S$^{2+}$) or S/S$^{+}$). 
RF input features include
He$^{2+}$/He$^{+}$, O$^{2+}$/O$^{+}$, S$^{2+}$/S$^{+}$, Ar$^{3+}$/Ar$^{2+}$, and Cl$^{3+}$/Cl$^{2+}$. 
We used the \texttt{RandomForestRegressor} 
in the Python library \textsc{Scikit-Learn} \citep{pedregosa2011scikit}. 
Objects with unobserved O$^{2+}$ were excluded due 
to large O/H uncertainties and challenges in accurately determining the O$^{2+}$/O$^{+}$ ratio. 
Where other ionic stages in input features were not seen, such as Ar$^{3+}$ or Cl$^{3+}$, a separate 
RF regressor predicted the ionic fraction ratios (e.g., Ar$^{3+}$/Ar$^{2+}$ and Cl$^{3+}$/Cl$^{2+}$) 
using He$^{2+}$/He$^{+}$ and O$^{2+}$/O$^{+}$ as input features.

An extensive photoionization model grid of PNe was computed as in \citet{delgado2014ionization}. 
We selected PNe photoionization models from an extended grid version with a larger effective 
temperature range for PNe central stars. These models were run with version c17.01 of 
\textsc{Cloudy} \citep{ferland20172017} and hold in the \texttt{3MdB\_17} database 
\citep{morisset2015virtual} under the reference \texttt{PNe\_2020}. From the $>700,000$ models we chose 
$\sim120,000$ that met realistic PN model selection criteria in \citet{delgado2014ionization} 
to train the RF regressor. The full sample was divided into a training set (75\% of the sample), 
a validation set (12.5\%), and a test set (12.5\%). The training set trained the RF regressor 
and the validation set tuned the hyperparameter $\alpha$ of 
the RF regressor to enhance efficiency and accuracy achieved via a 
grid-search cross-validation (CV) of $\alpha$ values (see Appendix~\ref{appendix:rf_params}).

Figure~\ref{fig:sh_oh_ML} displays the distribution of differences between the $12+\log$(S/H) 
values obtained using the icf(S) from literature-based ICF schemes and our ML approach for the 
PNe sample in HSK12+ and Paper~III. 
The median differences for these samples are $-0.04_{-0.24}^{+0.13}$ and $-0.0006^{+0.15}_{-0.15}$~dex. 
The ICF scheme in DMS14, applied in Paper~III, shows no systematic difference and smaller discrepancies 
to the ML values, unlike the KW01 used in HSK12. The ML comparison for HSK12+ (right panel of 
Figure~\ref{fig:sh_oh_ML}) has a left-skewed distribution, suggesting underestimation of S/H and, 
consequently, a slightly larger average sulfur deficit of $\sim$0.07~dex. The best-fit line parameters for 
ML S/H values are presented in Table~\ref{tab:fit_params} for both HSK12+ and Paper~III PNe samples. 
The higher $R^{2}$ and $r$ values indicate a better linear relationship, while the best-fit 
line for HSK12+ is closer to unity. Considering the sulfur anomaly typically exceeds 0.1~dex, 
the inaccuracies linked to the ICF schemes are likely not the only cause of the sulfur anomaly. 
The increased consistency between the ICF scheme in DMS14 and the ML results shows that the DMS14 scheme is 
more reliable, supporting the argument that their ICF scheme reduces the sulfur anomaly, 
as shown here.
\begin{figure}
    \centering
    \includegraphics[width=0.48\textwidth]{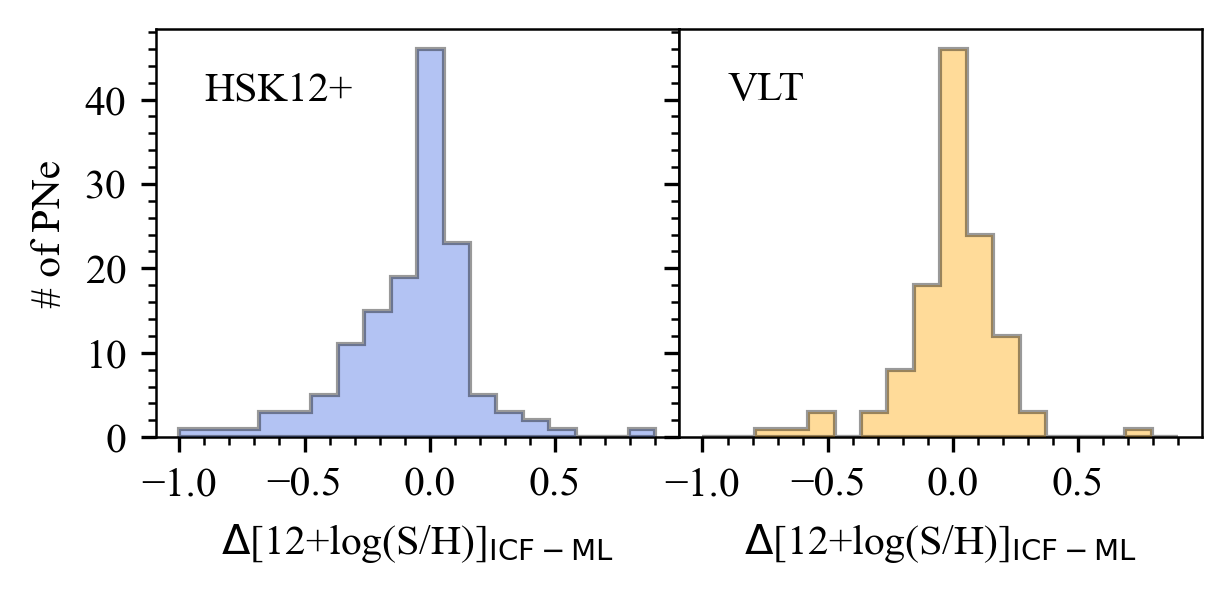}
    \caption{Differences between $12+\log$(S/H) values from literature ICF formulae and our machine 
    learning approach. The HSK12+ sample in the left panel adopted the KW01 ICF scheme, while the 
    right panel presents PNe from Paper~III using the DMS14 ICF scheme. The left-skewed distribution in 
    the HSK12+ panel suggests a contribution to the sulfur anomaly from the KW01 ICF scheme.}
    \label{fig:sh_oh_ML}
\end{figure}

\subsection{Mid-infrared observations of PNe}
\begin{figure}
    \centering
    \includegraphics[width=0.47\textwidth]{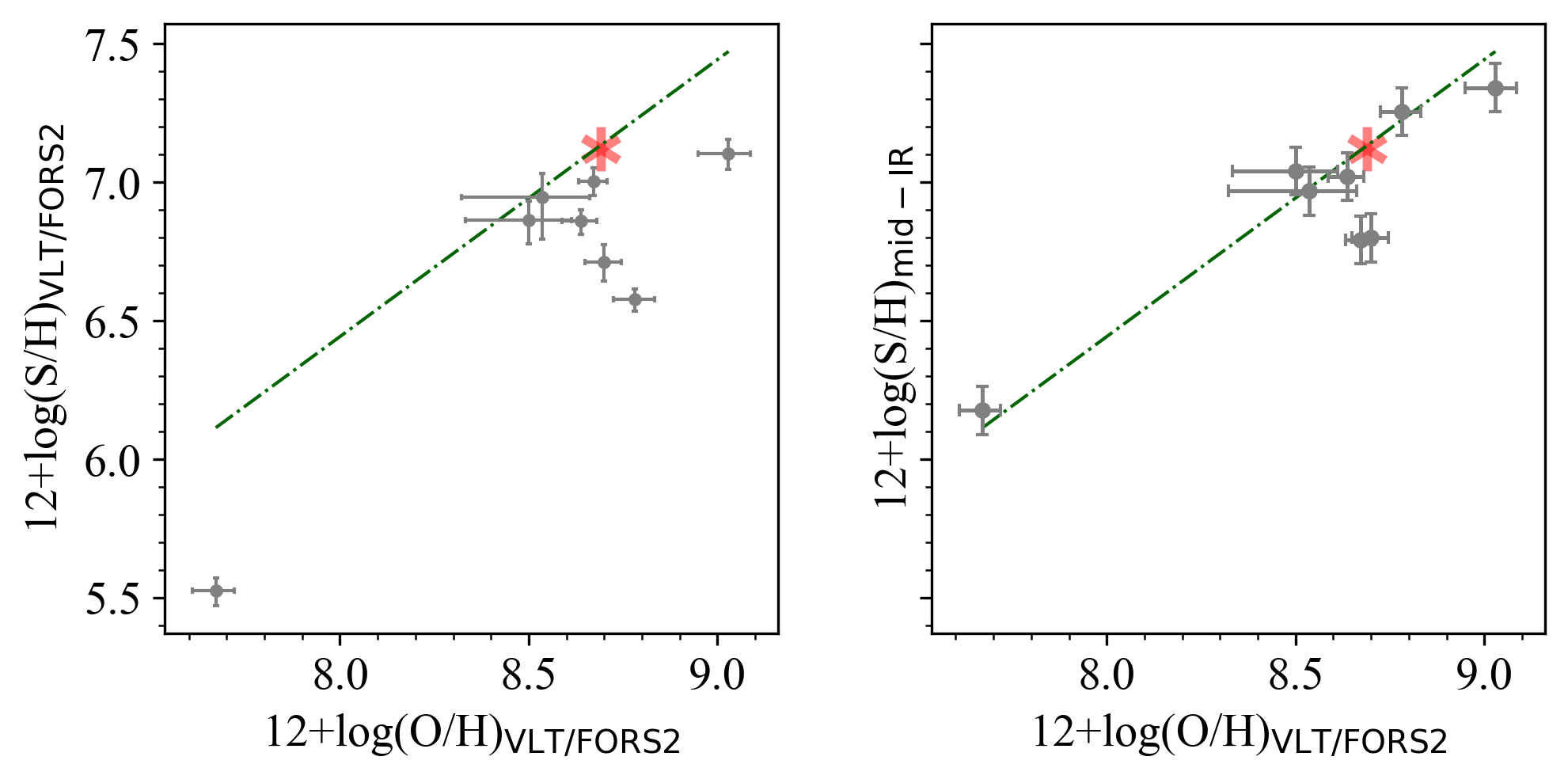}
    \caption{Comparison of S/H versus O/H for 8 bulge PNe in common with 
    \citet{pottasch2015abundances}. The left panel shows S/H values from optical 
    VLT observations in Paper~III. The right panel gives S/H values inferred 
    from PBS15 mid-infrared data. The dashed green line is the fit to 
    the H2BCG data; the red asterisk denotes the solar abundance. 
    For 6 out of 8 PNe in common the mid-infrared S/H values essentially 
    follow the expected trend line.}
    \label{fig:sul_abun_MIR}
\end{figure}

The S/H deficiency seen in Galactic PNe is 
suggested as due to underestimation of higher ionization stages of sulfur in 
photoionization models, e.g. \citet{henry2012curious, reichel2022recombination}. 
S/H values from mid-infrared \emph{Spitzer} 
spectra in \citet{pottasch2015abundances}, PBS15 hereafter, 
enables observation of the higher ionization stage S$^{3+}$ and are 
systematically higher than our results, with a median difference of 0.17~dex. 
The left panel of Fig.~\ref{fig:sul_abun_MIR} shows the VLT S/H versus O/H plot for 
eight bulge PNe in common with PBS15. 
The VLT results show the sulphur anomaly in terms of the H2BCG linear fit 
with an $R^{2}$ of 0.34. The PBS15 O/H values
are from optical observations but our VLT spectra are superior. 
The right panel shows good agreement between PBS15 sulfur abundances from mid-infrared 
spectra and oxygen abundances from our VLT spectra. We find $R^{2}=0.71$, 
consistent with the H2BCG results. A comparison of our sulfur ionic abundances and PBS15 
shows that the S$^{2+}$/H$^{+}$ mid-infrared values from [S~{\sc iii}] 
$\lambda$18.7~$\mu m$ and $\lambda$36.5~$\mu m$ in PBS15 are systematically higher 
than from the equivalent optical [S~{\sc iii}]~$\lambda$6312 line, 
with a median difference of 0.16~dex. The two outliers in the right hand plot in 
Fig.~\ref{fig:sul_abun_MIR} have no lower S$^{2+}$/H$^{+}$ values from mid-infrared 
spectra. The dashed green line is the H2BCG fit and the red asterisk is 
solar abundance. For 6/8 PNe in common the mid-infrared S/H values 
follow the expected trend line. Comparison of our icf(S) and S/(S$^{+}$+S$^{2+}$) 
or S/S$^{+}$ in PBS15 shows a median difference of $\sim$4\%, with underestimation 
$>$10\% in only two PNe. 

Hence, the sulfur anomaly is unlikely to arise 
from underestimation of higher sulfur ionization stages in the photoionization model and their 
ICFs. Optical measurement of [S~{\sc iii}] lines gives lower S$^{2+}$ abundances compared 
to their mid-infrared counterparts, if the 
values of S$^{2+}$ inferred from the mid-infrared are more accurate.

We highlight some issues associated with mixing the 
optical and mid-infrared data for chemical abundance derivation. 
First, Appendix.\ref{app:ne_ar_midIR} gives the same plots for neon and argon as for sulfur 
in Fig.~\ref{fig:sul_abun_MIR}. Our Paper~III VLT results show good
agreement with the H2BCG data, while the mid-infrared data yields systematically higher results 
by 0.29 and 0.33~dex for neon and argon. A similar issue with neon abundance was reported in 
\citet{dors2013optical} when studying star-forming regions. This was partly attributed 
to abundance variations across the nebulae. It is possible mid-infrared 
and optical data intrinsically provide different abundance estimates, with the mid-infrared data 
yielding higher estimates. This is supported by detailed PNe photoionization 
modelling in \citet{bandyopadhyay2021compact} and dusty PNe photoionization 
modelling in \citet{otsuka2017herschel}, where best-fit photoionization models for 
the optical emission lines and PNe physical conditions yielded lower fluxes 
of [S~{\sc iii}] $\lambda$18.7 $\mu$m and $\lambda$36.5~$\mu$m than the observed mid-infrared line fluxes.

\subsection{Underestimation of S\texorpdfstring{$^{2+}$}//H\texorpdfstring{$^{+}$} -from optical spectra}
To address the potential underestimation of S$^{2+}$ ionic abundances using optical spectra that 
could lead to a discrepancy between optical and mid-infrared observations, we re-examined our Paper~III S/H 
measurements. We considered several factors, including errors in interstellar correction at the red [S~{\sc 
iii}] lines from the use of a single, standard 
extinction law that may not be valid in the Galactic bulge 
\citep{gould2001method, udalski2003optical}, uncertainties in [S~{\sc iii}] line flux measurements 
and the varying spatial distribution of sulfur and oxygen in PNe.

To quantify the impact of interstellar extinction correction uncertainties, we recalculated 
S/H using emission line lists from Paper~III and compared two extinction laws 
introduced in \citet{howarth1983lmc} and \citet{cardelli1989relationship}, with that of 
\citet{fitzpatrick1999correcting} used in Paper~III. We showed good agreement between the S$^{2+}$/H$^{+}$ 
and S/O abundance ratios derived using the \citet{howarth1983lmc} and \citet{fitzpatrick1999correcting} 
extinction laws, with a median difference of $-0.001^{+0.004}_{-0.006}$~dex and $-0.00\pm0.007$~dex.  
The extinction law of \citet{cardelli1989relationship} yielded slightly lower values, 
indicating even lower results compared to the mid-infrared, with median differences in S$^{2+}$/H$^{+}$ 
and $\log$(S/O) of $-0.07^{+0.04}_{-0.06}$ and $-0.03\pm 0.03$~dex. We find the extinction 
correction impact on the sulfur anomaly is insignificant compared to the discrepancy between 
optical and mid-infrared S$^{2+}$ results and conclude that the extinction correction is not a 
contributing factor to the anomaly.

Our Paper~III VLT data covered a wavelength range up to 8500~\AA, which only allowed 
detection of the [S~{\sc iii}]~$\lambda$6312 line. We investigated whether lower signal-to-noise 
ratios for the [S~{\sc iii}]~$\lambda$6312 line might lead to underestimation of its line fluxes. 
Our analysis showed that the sulfur anomaly size had no correlation with 
measured raw fluxes of the [S~{\sc iii}]~$\lambda$6312 line, indicating that the sulfur 
deficiency is not due to weak line fluxes. 

The near-IR [S~{\sc iii}] $\lambda\lambda$9069, 9531 lines were not observed in our VLT spectra. The ionization structure of the PNe may introduce uncertainties in the S$^{2+}$ abundance determinations due the absence of electron temperatures derived from [S~{\sc iii}] line ratios, as S$^{2+}$ (22.34–34.79~eV) probes lower ionization zones than O$^{2+}$ (35.12-54.94~eV).
While the analysis of H~{\sc ii} regions utilising deep spectra in \cite{2013MNRAS.428.3660F} suggests no such systematic difference between T$_{\mathrm{e}}$([S~{\sc iii}]) and T$_{\mathrm{e}}$([O~{\sc iii}]), the ionising sources and the gas and dust components in H~{\sc ii} regions are considerably different from PNe. A comprehensive exploration of high-quality PNe data with the near-IR [S~{\sc iii}] line observations, where the telluric contamination can be satisfactorily removed, is imperative to examine how the ionization structures of PNe affect the sulfur deficit.

Abundances from long-slit spectroscopy only sample a portion of a PN often weighted towards high-ionization regions near the central star \citep{ali2016ifu}. Hence, we also consider literature on PNe integral field unit (IFU) spectroscopy. Such 2-D spectroscopy can provide global abundance data. 
Using ionic abundances of O and S derived from IFU observations, the sulfur anomaly persists for four PNe in \citet{ali2016ifu} and three PNe in \citet{garcia2022muse}. For most objects, the sulfur deficit reduced by 0.10 to 0.34 dex; however, two PNe show a greater $d$(S) by $\sim0.12$ dex using IFU observations. Thus, ionization stratification and spatial variation in sulfur and oxygen abundances within a PN may partly account for the observed sulfur anomaly.

To conclude, firstly HKS12 and Paper~III employ a non-biased ICF for sulfur, as shown by our
ML icf(S) values. Secondly, possible intrinsic differences between optical 
and mid-infrared spectroscopy results may lead to higher abundances derived from 
mid-infrared spectra. Thirdly, among the application 
of different extinction laws, uncertainties in  measuring weak [S~{\sc iii}]~$\lambda$6312 lines and any 
abundance inhomogeneities within the PNe, none appear to have a significant impact that could 
lead to underestimation of S$^{2+}$ ionic abundances. The PNe sulfur anomaly, though reduced in our work, 
remains a genuine effect but is unlikely to arise from underestimating 
sulfur ionization stages above S$^{2+}$.

\section{The sulfur deficit across different PNe samples and properties} 
\noindent
After confirming in Section~\ref{sec:confirm_SA} that the sulfur anomaly is a real, albeit 
much reduced  in PNe, it is likely that the varying PNe S/H deficiencies
in different Galactic regions, as discussed in Section~\ref{sec:res}, could 
indicate a progenitor population effect. Below we present an analysis of the sulfur deficit 
considering links to PN progenitor age/mass, dust chemistry and Galactic location. 
The results are summarised in Fig.~\ref{fig:SD_age_scatter_hist} and explored in terms of 
PNe progenitor age and sulfur deficit.

\subsection{PNe progenitor ages/masses}
We developed a basic PNe age estimator 
based on our high-quality VLT abundances as detailed in Appendix~\ref{appendix:pne_ages}. We 
differentiate between young ($t_{\star}<1$~Gyr) and old ($t_{\star}>7.5$~Gyr) PN populations by combining 
the Peimbert classes with an extant scheme in \citet{stanghellini2018galactic} that establish a link between 
the effectiveness of nitrogen production in PN progenitors and their ages/masses. We used this criteria to 
divide the HSK12+ disk sample and our VLT bulge sample into crude young and old progenitor star PNe 
populations. 

In Fig.~\ref{fig:SD_age_scatter_hist}, we give a combined plot for sulfur deficit, $d$(S), 
as a function of Galactic height in kpc; 12~+~$\log$(O/H) and 
histograms of $d$(S) values with bins of 0.11~dex for the HSK12+ disk PNe sample (top row) and 
VLT bulge PNe sample (bottom row). We color-coded the data according to these crude ages.

In the upper left panel of Fig.~\ref{fig:SD_age_scatter_hist}, the HSK12+ disk PNe show 
a broad spread in $d$(S) with a peak at $\sim0.4$. 
The old PNe population is restricted to a narrower range of higher $d$(S), while the younger 
population have a broader distribution from $1.0$ to $-0.6$~dex in $d$(S). Objects with no sulfur deficit 
primarily have young progenitors. In the lower left panel of Fig.~\ref{fig:SD_age_scatter_hist}, 
the bulge PNe sample has a narrower uni-modal $d$(S) distribution peaking at $\sim0.3$~dex. 
The old population is restricted to a narrower range of higher $d$(S) while the 
younger population dominates at lower $d$(S) values up to $-0.2$~dex. 

A two-sample Kolmogorov–Smirnov (KS) test \citep{smirnov1939estimation} 
using \texttt{scipy.stats.ks\_2samp} compared the $d$(S) distribution of the young and old 
populations. The KS p-values are 8E-4 for disk PNe and only 1E-5 for bulge PNe. Adopting a stringent 
significance level $\alpha$ of 0.001, we reject the null hypothesis that the $d$(S) of different age groups come from the same population. 

Given the crudeness of the age classification these results are indicative, but they imply the sulfur 
deficit is PN age related and hence to the initial progenitor mass.

\subsection{Galactic height from the mid-plane}
We use identifications of central stars of Galactic PNe from \emph{Gaia} EDR3 in \citet{chornay2021one} 
(CW21) and \citet{gonzalez2021planetary} (GSM21) and their parallax distances, to determine Galactic 
heights from the mid-plane. For Galactic disk PNe in HSK12+ and the VLT Galactic bulge PNe sample 
in Paper~III, there are 94 and 47 objects with \emph{Gaia} parallax distances. 
Preliminary analysis by \citet{parker2022preliminary} and \citet{tan2023morphologies}, indicated that 
the CW21 identifications are more reliable though up to 12\% of wrongly identified 
CSPN remain. We use the CW21 results when \emph{Gaia} distances were 
available in both studies. We calculated the Galactic Plane vertical distance ($z$) 
using \emph{Gaia} distances and Galactic ($l$,~$b$) coordinates.

The right panel of Fig.~\ref{fig:SD_age_scatter_hist} displays $d$(S) against $|z|$ in kpc. 
In the young disk PNe progenitor population, 62\% have $|z|<0.25$~kpc while the old population is more
evenly split. 
Indeed, the stellar population associated with the Galactic thin disk is expected to having 
younger members at lower Galactic scale heights \citep[e.g.][]{li2017evolution}. For the bulge 
sample the Galactic height issue is more complex probably due to the greater mixing of stellar 
populations. No correlation was found between $d$(S) and $|z|$ for the old PNe populations. 

\subsection{Metallicity indicator O/H}
\begin{figure}
    \centering
    \includegraphics[width=0.48\textwidth]{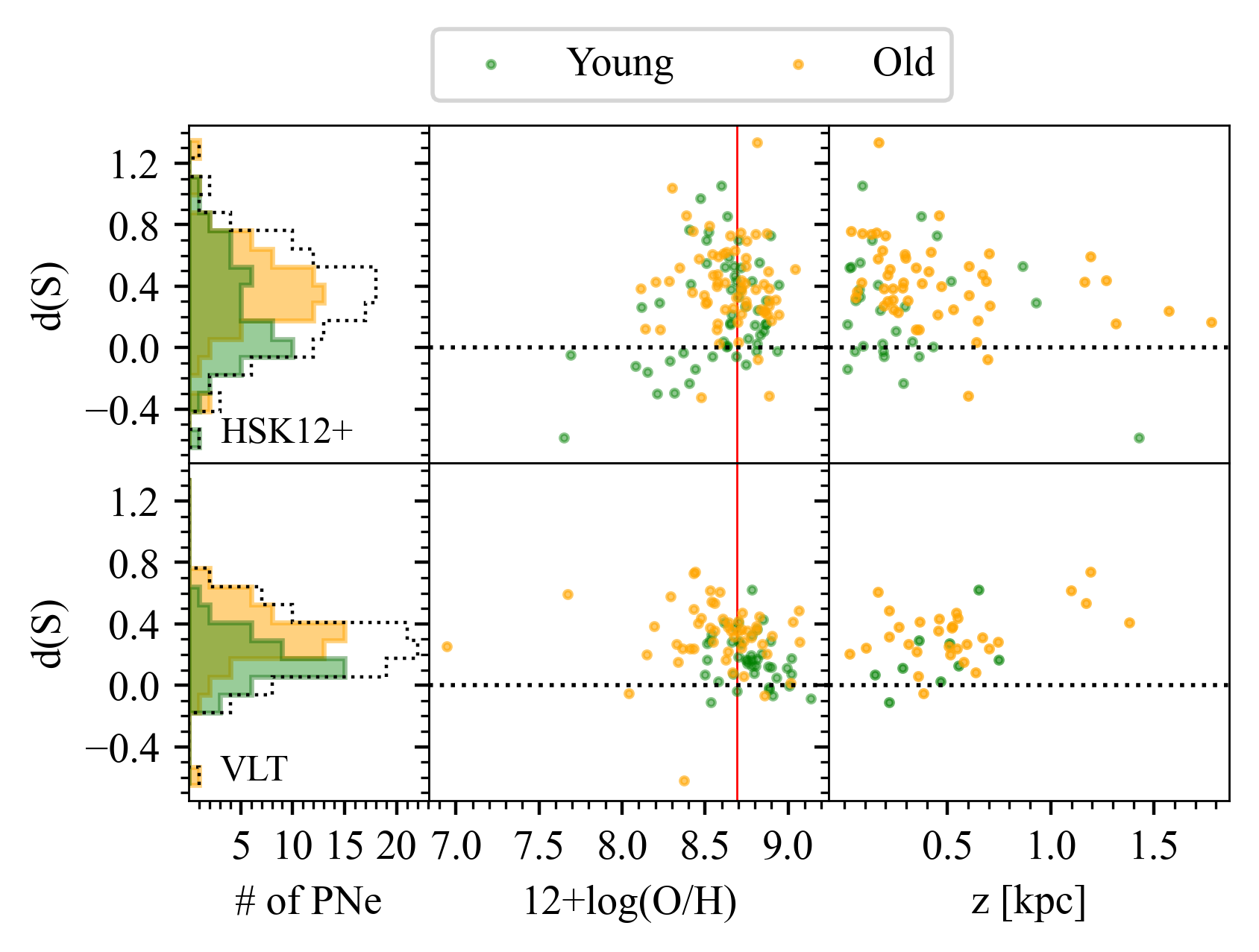}
    \caption{A combined figure for sulfur deficit $d$(S) on the vertical axis for the HSK12+ sample (top 
    row) and VLT sample (bottom row) against i) histograms of $d$(S) values with a bin size of 0.11~dex, ii) 
    12+$\log$(O/H) and iii) absolute height above the Galactic plane `$|z|$' in kpc. The Galactic disk and 
    bulge PNe samples are color-coded to their basic age group from criteria outlined in 
    Appendix~\ref{appendix:pne_ages}. Green indicates young and orange indicates old progenitor stars 
    according to the adopted dating scheme. The grey dashed line in the left panel outlines the overall 
    distribution. The middle panel red vertical line indicates solar oxygen abundance.}  
    \label{fig:SD_age_scatter_hist}
\end{figure}

The middle panel of Fig.~\ref{fig:SD_age_scatter_hist} depicts $d$(S) as a function of $12+\log$(O/H). 
For both the disk and bulge PNe the sulfur deficit exhibits a weak association with oxygen 
abundance. PNe with extreme sulfur deficits ($d$(S)$>0.5$~dex) exist mainly in the disk PNe with 
sub-solar to solar O/H values. Bulge PNe showing low or no sulfur deficit ($0.1<d$(S)~$<-0.1$~dex) 
have slightly sub-solar to super-solar O/H values and are from the young progenitor population. 
Some disk PNe with low or no sulfur deficit display sub-solar to super-solar O/H values. 
Several young disk objects have an overabundance of sulfur with sub-solar to extremely 
sub-solar O/H. 

At lower O/H values, $d$(S) increases as O/H increases, while at higher O/H, $d$(S) decreases as 
the O/H increases. The overall point colour distribution, particularly evident in the bulge sample suggests 
distinct mechanisms may drive the sulfur deficit in PNe, dominating in oxygen-poor and oxygen-rich nebulae.


\subsection{PN morphology}
Morphological classifications of disk and bulge PNe are from HASH and Paper~I. Most 
PNe were classified as elliptical, bipolar or round. For both disk and bulge PNe, the morphological 
classes, based on their average sulfur deficit values, could be arranged in decreasing deficit as round, 
elliptical and bipolar. There is no major difference between $d$(S) values for the morphological 
classes, and their averages agree to 1$\sigma$. 

\subsection{PNe dust chemistry}
AGB stars produce different dust grains based on their C/O abundance ratio. Possible gas phase 
sulfur depletion due to dust or molecule formation was investigated in \citet{henry2012curious}. 
They concluded it is unlikely to explain the sulfur anomaly as no correlation between $d$(S) and 
C/O ratios was seen. The formation of key sulfur-bearing compounds, MgS and FeS, depends on a 
C-rich (C/O~$>1$) environment in AGB stars \citep[e.g.][]{nuth1985laboratory, lodders1995origin}. 
However, to classify PNe as C-rich or O-rich (C/O~$<1$), observing the dust features in their 
infrared spectra is a more reliable way than measuring the C/O abundance ratios from optical 
emission lines. The latter could be affected by uncertainties and depletion of carbon or oxygen 
into dust grains \citep{delgado2015oxygen}. 

PNe dust features from \emph{Spitzer}
\citep{garcia2014chemical} show that 33\% of Galactic disk PNe and 8\% of bulge PNe have C-rich 
dust features. Such C-rich PNe generally have sub-solar O/H and a larger sulfur 
deficit (median $d$(S)$=$0.51~dex) while those with O-rich or C-rich and O-rich dust 
features (dual chemistry) scatter above and below the BCG data, with median $d$(S) 
values of 0.34 and 0.07~dex (see their Fig.~8).

This work suggests missing gas-phase sulfur in C-rich PNe could be due to efficient dust formation, 
while reduced sulfur deficit in O-rich objects may be due to sulfur depletion into 
molecules \citep[e.g. SO, SO2, H2S, and CS][]{omont1993s, bujarrabal1994molecular}
in C-rich and O-rich environments. Furthermore, \citet{delgado2015oxygen} found that 
C-rich PNe with sub-solar O/H were enriched in oxygen by $\sim0.3$~dex. Hence, the 
sulfur deficit in C-rich nebulae may be due to both dust and/or molecule depletion and oxygen 
enrichment, while gas-phase sulfur depletion into molecules leads to a smaller sulfur 
deficit in oxygen rich PNe. 

AGB star evolution can lead to transformation of low to intermediate mass stars (1.5-4$M_{\odot}$) 
from O-rich to C-rich stars through the efficient third dredge-up process \citep{karakas2014dawes}. 
However, low-mass stars ($\le1.5$$M_{\odot}$) and more massive stars ($>$~3-4$M_{\odot}$) which are 
classified as young PNe in this study, are expected to remain O-rich throughout their evolution.

\section{Summary}
Based on our robust abundances for a significant sample of Galactic bulge PNe, we report 
important new insights into the PNe sulfur anomaly while reducing its significance.  We rule out some 
previous suggestions of where it might 
originate and validate others that contribute to its reduced form. We reveal that it is the young and more massive 
progenitors of the current PNe population and their associated dust chemistry that show low or absent sulfur deficit. Our key findings are:

    * we show lockstep behaviour for sulfur versus both oxygen and argon abundances for the first time, 
    emphasising the role high-quality data plays
    
    * adopting the DMS14 ICF scheme reduces the sulfur anomaly as our machine learning approach shows
    
    * based on mid-infrared data for a small, overlapping PNe sample, we find the sulfur anomaly is unlikely 
    due to underestimation of higher sulfur ionization stages in the photoionization models and their 
    derived ICFs 
    
    * the impact of different extinction laws on the sulfur anomaly is insignificant 
    
    * the sulfur deficit is not due to weak sulfur line fluxes in the optical range. While IFU observations 
    show sampling whole PNe is important for an overall sulfur deficit determination it is also reduced for 
    some PNe

    * correlations of lower sulfur deficit with Galactic height above/below the mid-plane and their close 
    relation to PNe progenitor mass and chemistry are shown
    
    * the sulfur deficit is almost absent for PNe from higher mass, younger progenitors and those with 
    oxygen or dual dust chemistry related to progenitor mass. Larger 
    deficits for intermediate progenitor mass PNe that are carbon rich are seen, the solution is 
    likely there

\begin{acknowledgments}
{\it Acknowledgments}\\
We thank the anonymous referee for their prompt review and constructive comments that improved this manuscript. ST thanks HKU and QAP for an MPhil scholarship and a research assistantship. QAP thanks the Hong Kong 
Research Grants Council for GRF research grants 17326116, 17300417 and 17304520. We made use of NASA’s Astrophysics 
Data System; the 
SIMBAD database, operated at CDS, Strasbourg, France; Astropy, a community-developed core Python 
package for Astronomy \citep{robitaille2013astropy}; HASH, an online database at the Laboratory for Space 
Research at HKU federates available multi-wavelength imaging, spectroscopic and other data for all known 
Galactic PNe and is available at: \url{http://www.hashpn.space}.
\end{acknowledgments}

%






\bibliography{Bulge}{}

\begin{thebibliography}{}
\expandafter\ifx\csname natexlab\endcsname\relax\def\natexlab#1{#1}\fi
\providecommand{\url}[1]{\href{#1}{#1}}
\providecommand{\dodoi}[1]{doi:~\href{http://doi.org/#1}{\nolinkurl{#1}}}
\providecommand{\doeprint}[1]{\href{http://ascl.net/#1}{\nolinkurl{http://ascl.net/#1}}}
\providecommand{\doarXiv}[1]{\href{https://arxiv.org/abs/#1}{\nolinkurl{https://arxiv.org/abs/#1}}}

\bibitem[{Ali {et~al.}(2016)Ali, Dopita, Basurah, Amer, Alsulami, \&
  Alruhaili}]{ali2016ifu}
Ali, A., Dopita, M., Basurah, H., {et~al.} 2016, Monthly Notices of the Royal
  Astronomical Society, 462, 1393

\bibitem[{Bandyopadhyay {et~al.}(2021)Bandyopadhyay, Das, \&
  Mondal}]{bandyopadhyay2021compact}
Bandyopadhyay, R., Das, R., \& Mondal, S. 2021, Monthly Notices of the Royal
  Astronomical Society, 504, 816

\bibitem[{Breiman(2001)}]{breiman2001random}
Breiman, L. 2001, Machine learning, 45, 5

\bibitem[{Bujarrabal {et~al.}(1994)Bujarrabal, Fuente, \&
  Omont}]{bujarrabal1994molecular}
Bujarrabal, V., Fuente, A., \& Omont, A. 1994, Astronomy and Astrophysics, Vol.
  285, p. 247-271 (1994), 285, 247

\bibitem[{Calvet \& Peimbert(1983)}]{calvet1983bipolar}
Calvet, N., \& Peimbert, M. 1983, Revista Mexicana de Astronomia y Astrofisica,
  5, 319

\bibitem[{Cardelli {et~al.}(1989)Cardelli, Clayton, \&
  Mathis}]{cardelli1989relationship}
Cardelli, J.~A., Clayton, G.~C., \& Mathis, J.~S. 1989, Astrophysical Journal,
  Part 1 (ISSN 0004-637X), vol. 345, Oct. 1, 1989, p. 245-256., 345, 245

\bibitem[{Cavichia {et~al.}(2010)Cavichia, Costa, \&
  Maciel}]{cavichia2010planetary}
Cavichia, O., Costa, R., \& Maciel, W. 2010, Revista mexicana de
  astronom{\'\i}a y astrof{\'\i}sica, 46, 159

\bibitem[{Chornay \& Walton(2021)}]{chornay2021one}
Chornay, N., \& Walton, N. 2021, Astronomy \& Astrophysics, 656, A110

\bibitem[{Corradi \& Schwarz(1995)}]{corradi1995morphological}
Corradi, R.~L., \& Schwarz, H.~E. 1995, Astronomy and Astrophysics, Vol. 293,
  p. 871-888 (1995), 293, 871

\bibitem[{{Delgado-Inglada} {et~al.}(2014){Delgado-Inglada}, {Morisset}, \&
  {Stasi{\'n}ska}}]{2014MNRAS.440..536D}
{Delgado-Inglada}, G., {Morisset}, C., \& {Stasi{\'n}ska}, G. 2014, \mnras,
  440, 536, \dodoi{10.1093/mnras/stu341}

\bibitem[{Delgado-Inglada {et~al.}(2014)Delgado-Inglada, Morisset, \&
  Stasi{\'n}ska}]{delgado2014ionization}
Delgado-Inglada, G., Morisset, C., \& Stasi{\'n}ska, G. 2014, Monthly Notices
  of the Royal Astronomical Society, 440, 536

\bibitem[{Delgado-Inglada {et~al.}(2015)Delgado-Inglada, Rodr{\'\i}guez,
  Peimbert, Stasi{\'n}ska, \& Morisset}]{delgado2015oxygen}
Delgado-Inglada, G., Rodr{\'\i}guez, M., Peimbert, M., Stasi{\'n}ska, G., \&
  Morisset, C. 2015, Monthly Notices of the Royal Astronomical Society, 449,
  1797

\bibitem[{Dors~Jr {et~al.}(2013)Dors~Jr, H{\"a}gele, Cardaci,
  P{\'e}rez-Montero, Krabbe, V{\'\i}lchez, Sales, Riffel, \&
  Riffel}]{dors2013optical}
Dors~Jr, O.~L., H{\"a}gele, G.~F., Cardaci, M.~V., {et~al.} 2013, Monthly
  Notices of the Royal Astronomical Society, 432, 2512

\bibitem[{Ferland {et~al.}(2017)Ferland, Chatzikos, Guzm{\'a}n, Lykins,
  Van~Hoof, Williams, Abel, Badnell, Keenan, Porter,
  {et~al.}}]{ferland20172017}
Ferland, G., Chatzikos, M., Guzm{\'a}n, F., {et~al.} 2017, Revista mexicana de
  astronom{\'\i}a y astrof{\'\i}sica, 53

\bibitem[{Fitzpatrick(1999)}]{fitzpatrick1999correcting}
Fitzpatrick, E.~L. 1999, Publications of the Astronomical Society of the
  Pacific, 111, 63

\bibitem[{{Freeman} {et~al.}(2013){Freeman}, {Ness}, {Wylie-de-Boer},
  {Athanassoula}, {Bland-Hawthorn}, {Asplund}, {Lewis}, {Yong}, {Lane}, {Kiss},
  \& {Ibata}}]{2013MNRAS.428.3660F}
{Freeman}, K., {Ness}, M., {Wylie-de-Boer}, E., {et~al.} 2013, \mnras, 428,
  3660, \dodoi{10.1093/mnras/sts305}

\bibitem[{Garc{\'\i}a-Hern{\'a}ndez \& G{\'o}rny(2014)}]{garcia2014chemical}
Garc{\'\i}a-Hern{\'a}ndez, D., \& G{\'o}rny, S. 2014, Astronomy \&
  Astrophysics, 567, A12

\bibitem[{{Garc{\'\i}a-Hern{\'a}ndez} \&
  {G{\'o}rny}(2014)}]{2014A&A...567A..12G}
{Garc{\'\i}a-Hern{\'a}ndez}, D.~A., \& {G{\'o}rny}, S.~K. 2014, \aap, 567, A12,
  \dodoi{10.1051/0004-6361/201423620}

\bibitem[{Garc{\'\i}a-Rojas {et~al.}(2022)Garc{\'\i}a-Rojas, Morisset, Jones,
  Wesson, Boffin, Monteiro, Corradi, \& Rodr{\'\i}guez-Gil}]{garcia2022muse}
Garc{\'\i}a-Rojas, J., Morisset, C., Jones, D., {et~al.} 2022, Monthly Notices
  of the Royal Astronomical Society, 510, 5444

\bibitem[{Gonz{\'a}lez-Santamar{\'\i}a
  {et~al.}(2021)Gonz{\'a}lez-Santamar{\'\i}a, Manteiga, Manchado, Ulla,
  Dafonte, \& Varela}]{gonzalez2021planetary}
Gonz{\'a}lez-Santamar{\'\i}a, I., Manteiga, M., Manchado, A., {et~al.} 2021,
  Astronomy \& Astrophysics, 656, A51

\bibitem[{Gould {et~al.}(2001)Gould, Stutz, \& Frogel}]{gould2001method}
Gould, A., Stutz, A., \& Frogel, J.~A. 2001, The Astrophysical Journal, 547,
  590

\bibitem[{Henry {et~al.}(2010)Henry, Kwitter, Jaskot, Balick, Morrison, \&
  Milingo}]{henry2010abundances}
Henry, R., Kwitter, K.~B., Jaskot, A.~E., {et~al.} 2010, The Astrophysical
  Journal, 724, 748

\bibitem[{Henry {et~al.}(2004)Henry, Kwitter, \& Balick}]{henry2004sulfur}
Henry, R.~B., Kwitter, K., \& Balick, B. 2004, The Astronomical Journal, 127,
  2284

\bibitem[{Henry {et~al.}(2012)Henry, Speck, Karakas, Ferland, \&
  Maguire}]{henry2012curious}
Henry, R.~B., Speck, A., Karakas, A.~I., Ferland, G.~J., \& Maguire, M. 2012,
  The Astrophysical Journal, 749, 61

\bibitem[{Howarth(1983)}]{howarth1983lmc}
Howarth, I.~D. 1983, Monthly Notices of the Royal Astronomical Society, 203,
  301

\bibitem[{Izotov \& Thuan(1999)}]{izotov1999heavy}
Izotov, Y.~I., \& Thuan, T.~X. 1999, The Astrophysical Journal, 511, 639

\bibitem[{Karakas \& Lattanzio(2014)}]{karakas2014dawes}
Karakas, A.~I., \& Lattanzio, J.~C. 2014, Publications of the Astronomical
  Society of Australia, 31, e030

\bibitem[{Kennicutt~Jr {et~al.}(2003)Kennicutt~Jr, Bresolin, \&
  Garnett}]{kennicutt2003composition}
Kennicutt~Jr, R.~C., Bresolin, F., \& Garnett, D.~R. 2003, The Astrophysical
  Journal, 591, 801

\bibitem[{Kingsburgh \& Barlow(1994)}]{kingsburgh1994elemental}
Kingsburgh, R.~L., \& Barlow, M. 1994, Monthly Notices of the Royal
  Astronomical Society, 271, 257

\bibitem[{Kwitter \& Henry(2001)}]{kwitter2001sulfur}
Kwitter, K., \& Henry, R. 2001, The Astrophysical Journal, 562, 804

\bibitem[{Li \& Zhao(2017)}]{li2017evolution}
Li, C., \& Zhao, G. 2017, The Astrophysical Journal, 850, 25

\bibitem[{Lodders \& Fegley~Jr(1995)}]{lodders1995origin}
Lodders, K., \& Fegley~Jr, B. 1995, Meteoritics, 30, 661

\bibitem[{{Mathis}(1985)}]{1985ApJ...291..247M}
{Mathis}, J.~S. 1985, \apj, 291, 247, \dodoi{10.1086/163063}

\bibitem[{Milingo {et~al.}(2010)Milingo, Kwitter, Henry, \&
  Souza}]{milingo2010alpha}
Milingo, J., Kwitter, K., Henry, R., \& Souza, S. 2010, The Astrophysical
  Journal, 711, 619

\bibitem[{Morisset {et~al.}(2015)Morisset, Delgado-Inglada, \&
  Flores-Fajardo}]{morisset2015virtual}
Morisset, C., Delgado-Inglada, G., \& Flores-Fajardo, N. 2015, Revista mexicana
  de astronom{\'\i}a y astrof{\'\i}sica, 51, 101

\bibitem[{Nuth {et~al.}(1985)Nuth, Moseley, Silverberg, Goebel, \&
  Moore}]{nuth1985laboratory}
Nuth, J.~A., Moseley, S.~H., Silverberg, R.~F., Goebel, J.~H., \& Moore, W.~J.
  1985, Astrophysical Journal, Part 2-Letters to the Editor (ISSN 0004-637X),
  vol. 290, March 1, 1985, p. L41-L43., 290, L41

\bibitem[{Omont {et~al.}(1993)Omont, Lucas, Morris, \& Guilloteau}]{omont1993s}
Omont, A., Lucas, R., Morris, M., \& Guilloteau, S. 1993, Astronomy and
  Astrophysics (ISSN 0004-6361), vol. 267, no. 2, p. 490-514., 267, 490

\bibitem[{Otsuka {et~al.}(2017)Otsuka, Ueta, Van~Hoof, Sahai, Aleman, Zijlstra,
  Chu, Villaver, Leal-Ferreira, Kastner, {et~al.}}]{otsuka2017herschel}
Otsuka, M., Ueta, T., Van~Hoof, P.~A., {et~al.} 2017, The Astrophysical Journal
  Supplement Series, 231, 22

\bibitem[{Parker {et~al.}(2022)Parker, Xiang, \&
  Ritter}]{parker2022preliminary}
Parker, Q.~A., Xiang, Z., \& Ritter, A. 2022, Galaxies, 10, 32

\bibitem[{Pedregosa {et~al.}(2011)Pedregosa, Varoquaux, Gramfort, Michel,
  Thirion, Grisel, Blondel, Prettenhofer, Weiss, Dubourg,
  {et~al.}}]{pedregosa2011scikit}
Pedregosa, F., Varoquaux, G., Gramfort, A., {et~al.} 2011, the Journal of
  machine Learning research, 12, 2825

\bibitem[{Peimbert(1978)}]{peimbert1978chemical}
Peimbert, M. 1978, in Symposium-International Astronomical Union, Vol.~76,
  Cambridge University Press, 215--224

\bibitem[{Peimbert \& Serrano(1980)}]{peimbert1980helium}
Peimbert, M., \& Serrano, A. 1980, Revista Mexicana de Astronomia y
  Astrofisica, vol. 5, Apr. 1980, p. 9-18., 5, 9

\bibitem[{Peimbert \& Torres-Peimbert(1983)}]{peimbert1983type}
Peimbert, M., \& Torres-Peimbert, S. 1983, in Symposium-International
  Astronomical Union, Vol. 103, Cambridge University Press, 233--242

\bibitem[{Phillips(2005)}]{phillips2005mean}
Phillips, J. 2005, Monthly Notices of the Royal Astronomical Society, 361, 283

\bibitem[{Pottasch \& Bernard-Salas(2015)}]{pottasch2015abundances}
Pottasch, S., \& Bernard-Salas, J. 2015, Astronomy \& Astrophysics, 583, A71

\bibitem[{Reichel {et~al.}(2022)Reichel, Kimeswenger, van Hoof, Zijlstra,
  Barr{\'\i}a, Hajduk, Van~de Steene, \& Tafoya}]{reichel2022recombination}
Reichel, M., Kimeswenger, S., van Hoof, P.~A., {et~al.} 2022, The Astrophysical
  Journal, 939, 103

\bibitem[{Robitaille {et~al.}(2013)Robitaille, Tollerud, Greenfield,
  Droettboom, Bray, Aldcroft, Davis, Ginsburg, Price-Whelan, Kerzendorf,
  {et~al.}}]{robitaille2013astropy}
Robitaille, T.~P., Tollerud, E.~J., Greenfield, P., {et~al.} 2013, Astronomy \&
  Astrophysics, 558, A33

\bibitem[{{Sabin} {et~al.}(2022){Sabin}, {G{\'o}mez-Llanos}, {Morisset},
  {G{\'o}mez-Gonz{\'a}lez}, {Guerrero}, {Todt}, \&
  {Fang}}]{2022MNRAS.511....1S}
{Sabin}, L., {G{\'o}mez-Llanos}, V., {Morisset}, C., {et~al.} 2022, \mnras,
  511, 1, \dodoi{10.1093/mnras/stab3649}

\bibitem[{{Shingles} \& {Karakas}(2013)}]{2013MNRAS.431.2861S}
{Shingles}, L.~J., \& {Karakas}, A.~I. 2013, \mnras, 431, 2861,
  \dodoi{10.1093/mnras/stt386}

\bibitem[{Smirnov(1939)}]{smirnov1939estimation}
Smirnov, N.~V. 1939, Bull. Math. Univ. Moscou, 2, 3

\bibitem[{Soderblom(2010)}]{soderblom2010ages}
Soderblom, D.~R. 2010, Annual Review of Astronomy and Astrophysics, 48, 581

\bibitem[{Stanghellini \& Haywood(2018)}]{stanghellini2018galactic}
Stanghellini, L., \& Haywood, M. 2018, The Astrophysical Journal, 862, 45

\bibitem[{Stanway \& Eldridge(2018)}]{stanway2018re}
Stanway, E.~R., \& Eldridge, J. 2018, Monthly Notices of the Royal Astronomical
  Society, 479, 75

\bibitem[{Tan {et~al.}(2023)Tan, Parker, Zijlstra, \&
  Ritter}]{tan2023morphologies}
Tan, S., Parker, Q.~A., Zijlstra, A., \& Ritter, A. 2023, Monthly Notices of
  the Royal Astronomical Society, 519, 1049

\bibitem[{{Tan} {et~al.}(2024){Tan}, {Parker}, {Zijlstra}, \&
  {Rees}}]{tan2023catalogue}
{Tan}, S., {Parker}, Q.~A., {Zijlstra}, A.~A., \& {Rees}, B. 2024, \mnras, 527,
  6363, \dodoi{10.1093/mnras/stad3496}

\bibitem[{Udalski(2003)}]{udalski2003optical}
Udalski, A. 2003, The Astrophysical Journal, 590, 284

\bibitem[{Ventura {et~al.}(2016)Ventura, Stanghellini, Di~Criscienzo,
  Garc{\'\i}a-Hern{\'a}ndez, \& Dell'Agli}]{ventura2016planetary}
Ventura, P., Stanghellini, L., Di~Criscienzo, M., Garc{\'\i}a-Hern{\'a}ndez,
  D., \& Dell'Agli, F. 2016, Monthly Notices of the Royal Astronomical Society,
  460, 3940

\end{thebibliography}
\bibliographystyle{aasjournal}

\appendix
\setcounter{figure}{0}
\setcounter{table}{0}

\section{Comparison of Ne/H and Ar/H versus
O/H relationships using optical and mid-
infrared observation}
\noindent
We present additional figures that give a comparison of Ne/H and Ar/H with O/H and with Ne/H and Ar/H derived from both the optical VLT data from Paper~III and the mid-infrared observations presented in PBS15.
\label{app:ne_ar_midIR}
\renewcommand{\thefigure}{A1}
\begin{figure}[ht]
    \centering
    \includegraphics[width=0.47\textwidth]{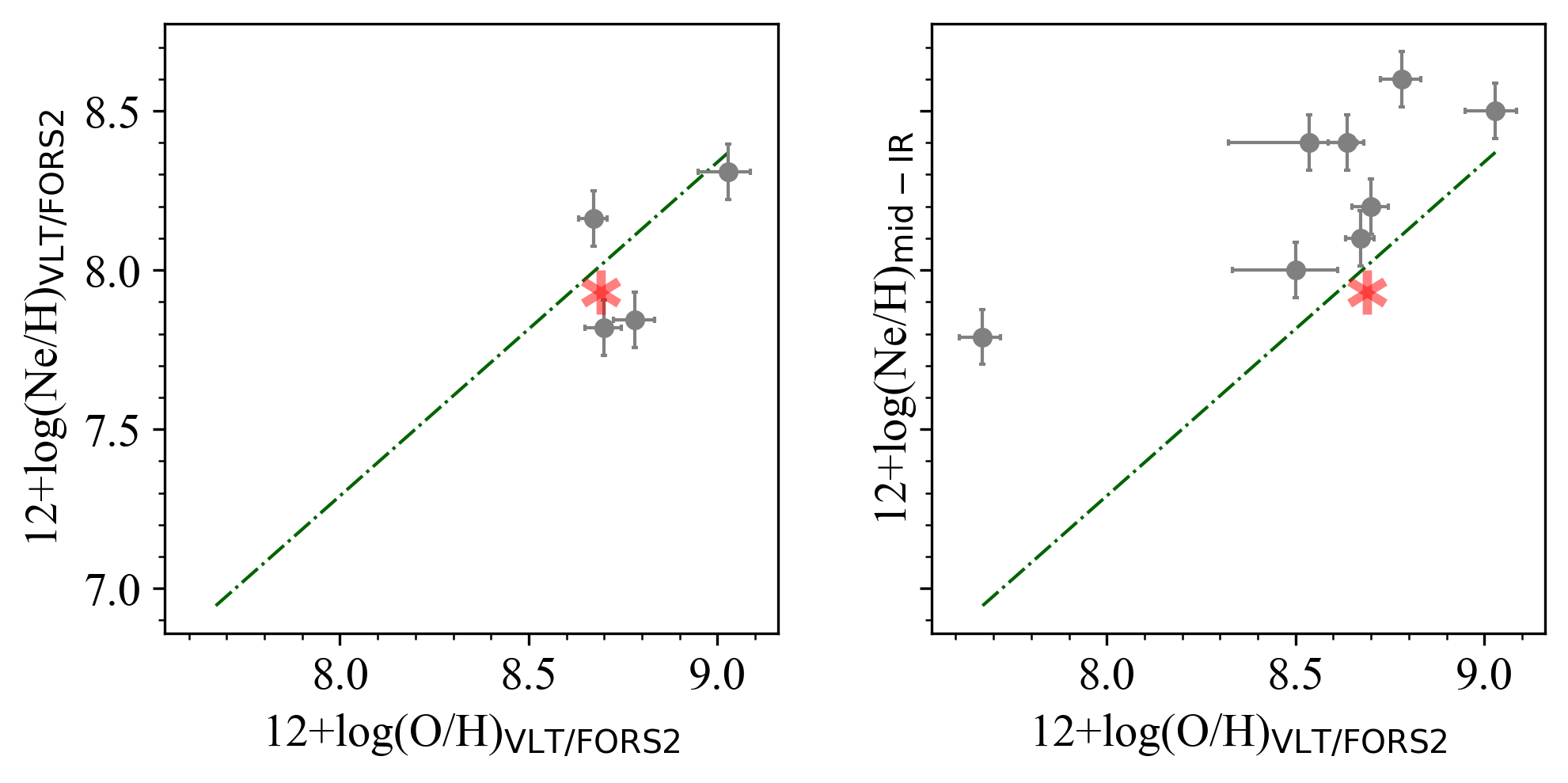}
    \includegraphics[width=0.47\textwidth]{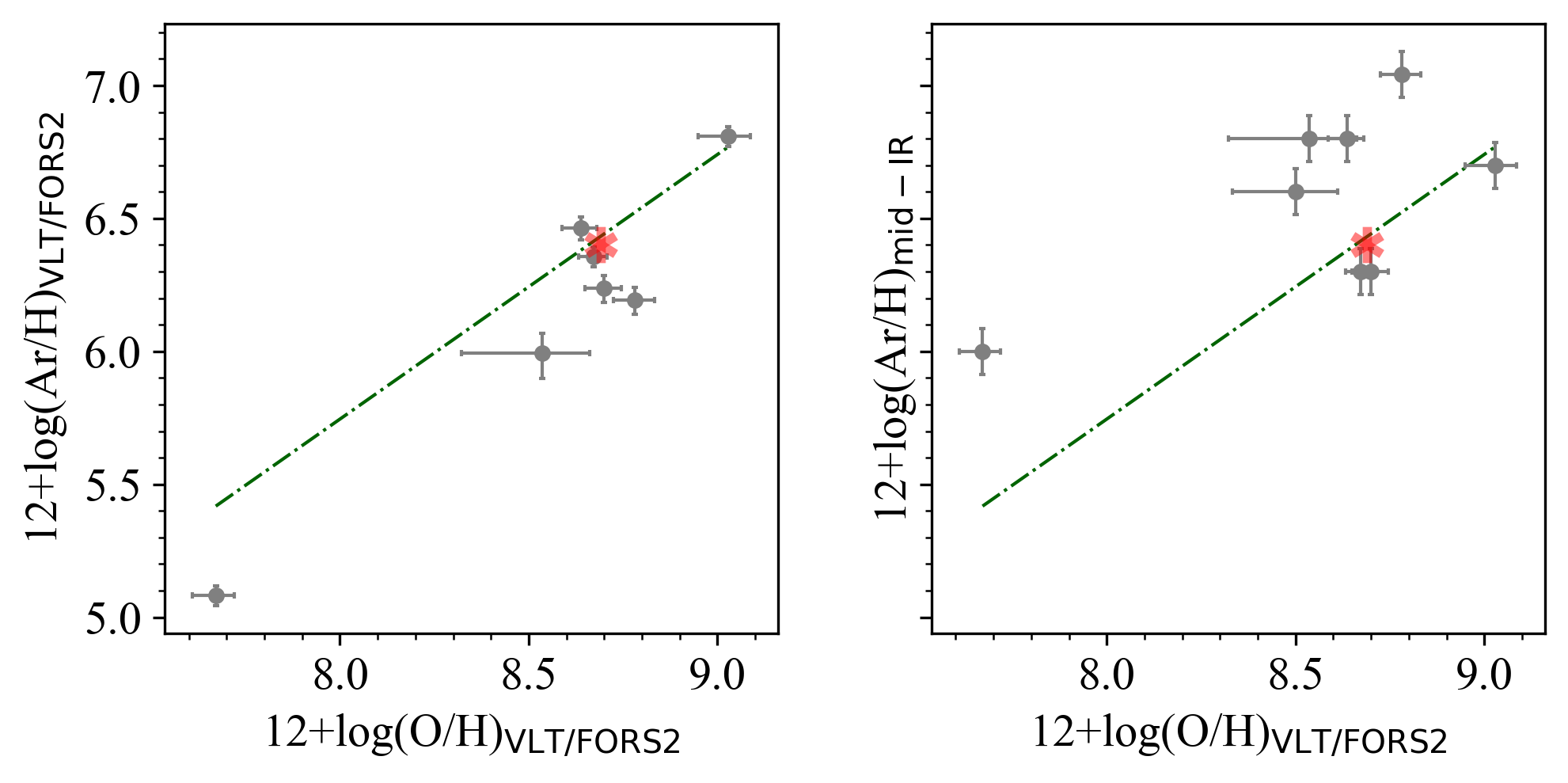}
    \caption{Comparison of Ne/H and Ar/H versus O/H relationships using optical VLT 
    observations and mid-infrared observations in PBS15, with symbols defined in 
    Fig.~\ref{fig:sul_abun_MIR}.}
    \label{fig:ne_ar_abun_MIR}
\end{figure}

\section{Selection of hyperparameters and performance evaluation for the random forest regressor}
\label{appendix:rf_params}
\noindent

The RF regressors, used to predict either the ionic fraction or the icf(S), all attained an $R^{2}$ of 0.99, 
indicating a reliable regression. For ICF values for sulfur, icf(S), used when S$^{2+}$ is observed, i.e. icf(S)$=$S/(S$^{+}$+S$^{2+}$), the fractional differences between the predicted and actual test sample 
values are generally $<5\%$. Only 0.9\% of cases exceed 15\% fractional differences (primarily for icf(S) 
values from 2.5 to 15). The median fractional difference is near zero (0.0002). This suggests no systematic 
bias between the predicted and true icf(S) values of the test set. For the icf(S) values 
used when only S$^{+}$ is observed, icf(S)$=$S/S$^{+}$, predicted values exhibit 
slightly larger deviations from test sample actual values. Nonetheless, the fractional difference is $<15\%$ 
for  $\sim$92\% of  cases, with a median fractional difference of 0.005. This indicates no significant 
systematic differences between the predicted and true icf(S) values of the test set.

To enhance the performance of the RF regressor, which was used for predicting either icf(S) or unobserved ionic fractions, we compared different sets of hyperparameters using the k-fold cross-validation (CV) on the validation set. The hyperparameters tuned for RF include the number of trees, (\texttt{n\_estimators}), the maximum depth of each tree, (\texttt{max\_depth}), the number of features allowed to make the best split while building the tree, (\texttt{max\_features}), minimum number of samples to split an internal node (\texttt{min\_samples\_split}), minimum number of samples to form a leaf node (\texttt{min\_samples\_leaf}) and whether bootstrap samples are used when building trees (\texttt{bootstrap}). The default values were used for the remaining hyperparameters. 
\subsection{RF regressor for icf(S)}
For icf(S), the RF model generates highly accurate predictions for both S/(S$^{2+}$+S$^{+}$) and S/S$^{+}$, with $R^{2}$ values exceeds 0.99. We presents the fractional uncertainties in predicted icf(S) with respect to the true values of the test set as a function of true icf(S) values on a log scale in Fig.\ref{fig:ML_icfS} for icf(S)$ = $S/(S$^{2+}$+S$^{+}$) (top panel) and icf(S)$ = $S/S$^{+}$ (bottom  panel). Both RF regressors exhibit no systematical discrepancies from the actual values, with most of the predicted values having a fractional uncertainty less than 5\% and 10\%, respectively. Although, the upper panel of Fig.\ref{fig:ML_icfS} reveals slightly higher predicted values than the actual values for intermediate S/(S$^{2+}$ + S$^{+}$) values ($\sim 1.26$ to 3.16), the median fractional uncertainty within this range is 0.01~dex. Thus, no systematical bias in predicted values presents for this range.
\renewcommand{\thetable}{B1}
\setlength{\tabcolsep}{2pt}
\begin{table}[ht]
    \centering
    \begin{tabular}{llcc}
\hline \hline
\multicolumn{1}{l}{\multirow{2}{*}{Parameter}} & \multicolumn{1}{l}{\multirow{2}{*}{Grid}} & \multicolumn{2}{c}{  Optimal parameter  }             
\\ \cline{3-4} \multicolumn{1}{c}{}     & \multicolumn{1}{c}{}   & S/(S$^{2+}$+S$^{+}$)       & S/S$^{+}$      
\\ \hline  n\_estimators      &  {[}100, 600{]}                           & 500           & 100       \\
min\_samples\_split                            & {[}2, 7{]}            & 3             & 3           \\
min\_samples\_leaf                             & {[}1, 10{]}           & 2             & 1          \\
max\_depth                                     & {[}5, 50{]}          & 13            & 50       \\
max\_features & {[}sqrt, auto{]}    & sqrt         & sqrt  \\
bootstrap                                      & {[}True, False{]}    & False         & True     \\ \hline
\end{tabular}
    \caption{Hyperparameter adjustments for icf(S) RF regressors.}
    \label{tab:rf_hyperparameter}
\end{table}

Each model's performance was evaluated with a 5-fold CV and quantified by the mean squared error (MSE) between the predictions and true values of the validation set. We used \texttt{RandomizedSearchCV} from \textsc{scikit-learn} to randomly select a subset of 50 among all combinations of $\alpha$ values. The best parameter selections are presented in Table~\ref{tab:rf_hyperparameter}. The optimal set of $\alpha$ values, which minimised the MSE, was chosen to train the final model. This model was then trained on the training set and applied to the test set.
\renewcommand{\thefigure}{B1}
\begin{figure}
    \centering
    \includegraphics[width=0.40\textwidth]{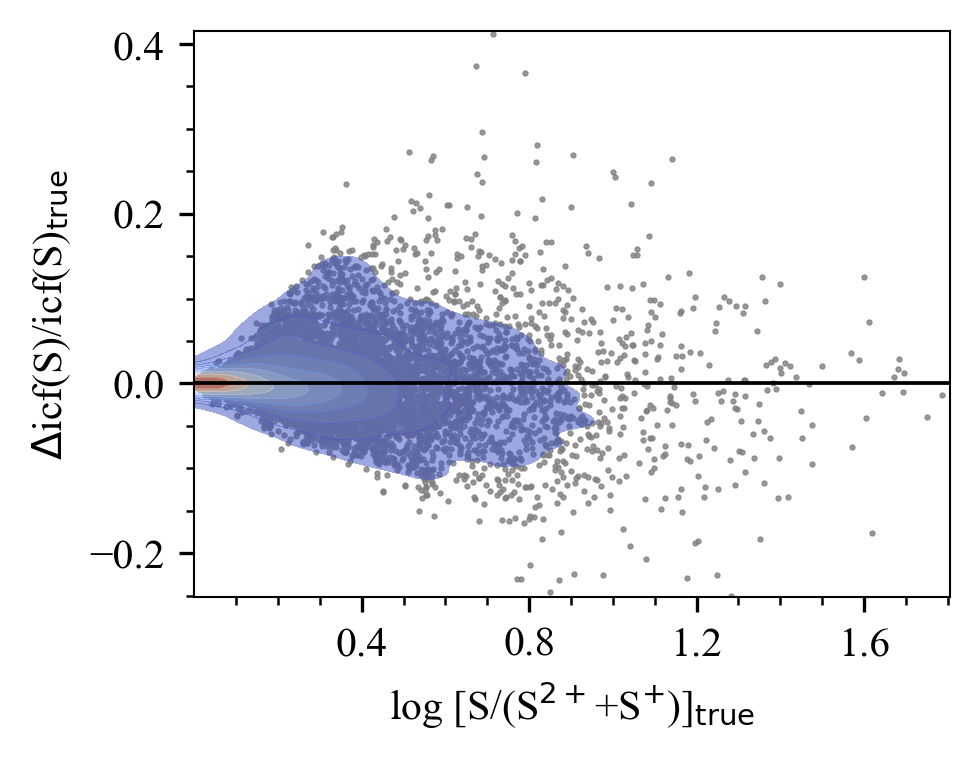}\hspace{0.7cm}
     \includegraphics[width=0.40\textwidth]{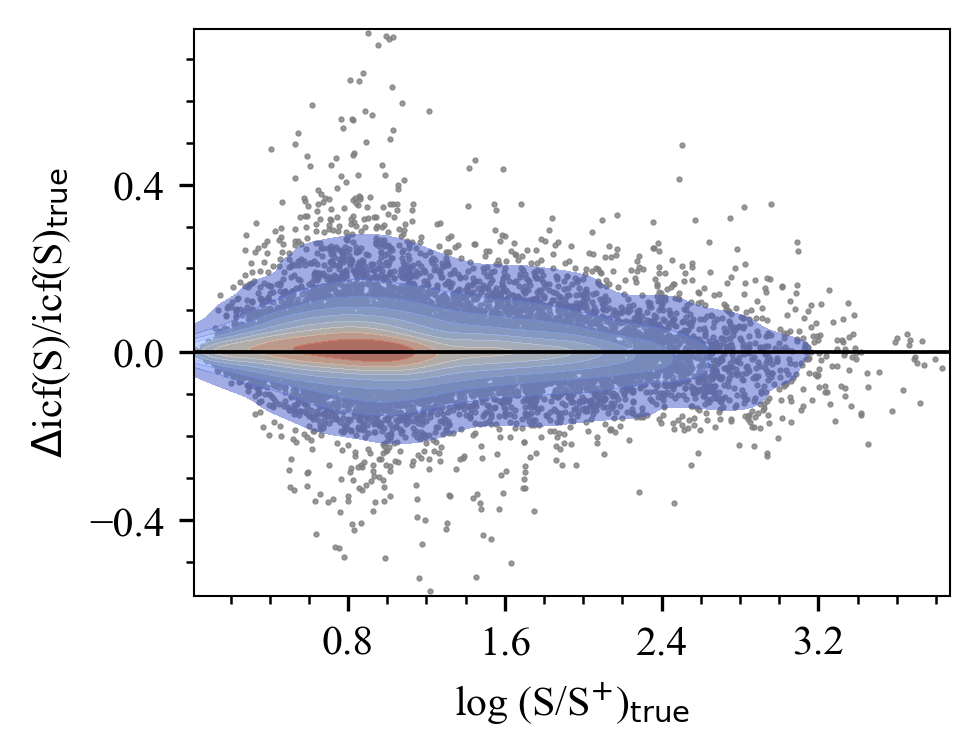}
    \caption{Fractional uncertainty in predicted icf(S) from the RF regressor as a function of log-scale true icf(S) for the test set. The contours illustrate the number density based on kernel density estimation. The black solid line denotes zero fractional uncertainty, with a high concentration of data points along this line indicating no systematic discrepancy between the predicted icf(S) from the RF regressor and the true values from the test set.}
    \label{fig:ML_icfS}
\end{figure}

\subsection{RF regressor for Ar and Cl ionic fractions }
We used two separated RF regressors to predict the unobserved ionic fraction, Ar$^{3+}$/Ar$^{2+}$, Cl$^{3+}$/Cl$^{2+}$, which are then used as input features to predict icf(S). Similarly, the hyperparameters were tuned through a 5-fold CV. We used \texttt{GridSearchCV} from \textsc{scikit-learn} to test all possible combinations of $\alpha$ values. The parameter grids and best parameter selections are presented in Table.\ref{tab:rf_hyperparameter_ion}. Using He$^{2+}$/He$^{+}$ and O$^{2+}$/O$^{+}$ as the input features, the RF models to predict  Ar$^{3+}$/Ar$^{2+}$ and Cl$^{3+}$/Cl$^{2+}$ achieved $R^{2+}$ values of 0.99. 
\renewcommand{\thetable}{B2}
\begin{table}
    \centering
    \begin{tabular}{llcc}
\hline \hline
\multicolumn{1}{l}{\multirow{2}{*}{Parameter}} & \multicolumn{1}{l}{\multirow{2}{*}{Grid}} & \multicolumn{2}{c}{Optimal parameter}        
\\ \cline{3-4} \multicolumn{1}{c}{}     & \multicolumn{1}{c}{}   & Ar$^{3+}$/Ar$^{2+}$    & Cl$^{3+}$/Cl$^{2+}$    
\\ \hline
n\_estimators   & {[}500, 2E3, 2E3{]}   & 2E3     & 1E3        \\
min\_samples\_split   & {[}2, 4, 6{]}   & 2   & 6   \\
min\_samples\_leaf    & {[}1, 5{]}   & 1   & 1          \\
max\_depth   & {[}50, 75, 100{]}     & 75     & 100         \\
max\_features & {[}sqrt, auto{]}    & auto         & auto  \\
bootstrap   & {[}True, False{]}   & True     & True       \\ \hline
\end{tabular}
    \caption{Hyperparameter adjustments for Ar$^{3+}$/Ar$^{2+}$ and  Cl$^{3+}$/Cl$^{2+}$ RF regressors.}
    \label{tab:rf_hyperparameter_ion}
\end{table}

The fractional uncertainties in  Ar$^{3+}$/Ar$^{2+}$ and Cl$^{3+}$/Cl$^{2+}$ as a function of the ionic fractions on a log scale is given in Fig.\ref{fig:ML_icfS_ionfrac}. The predicted values show larger deviation when Ar$^{2+}$ or Cl$^{2+}$ is the dominant ionization stage of Ar or Cl. Still, the predicted values agree with the true values of the test set within 50\% for 89\% and 90\% of the cases.  
\renewcommand{\thefigure}{B2}
\begin{figure}[ht]
    \centering
    \includegraphics[width=0.40\textwidth]{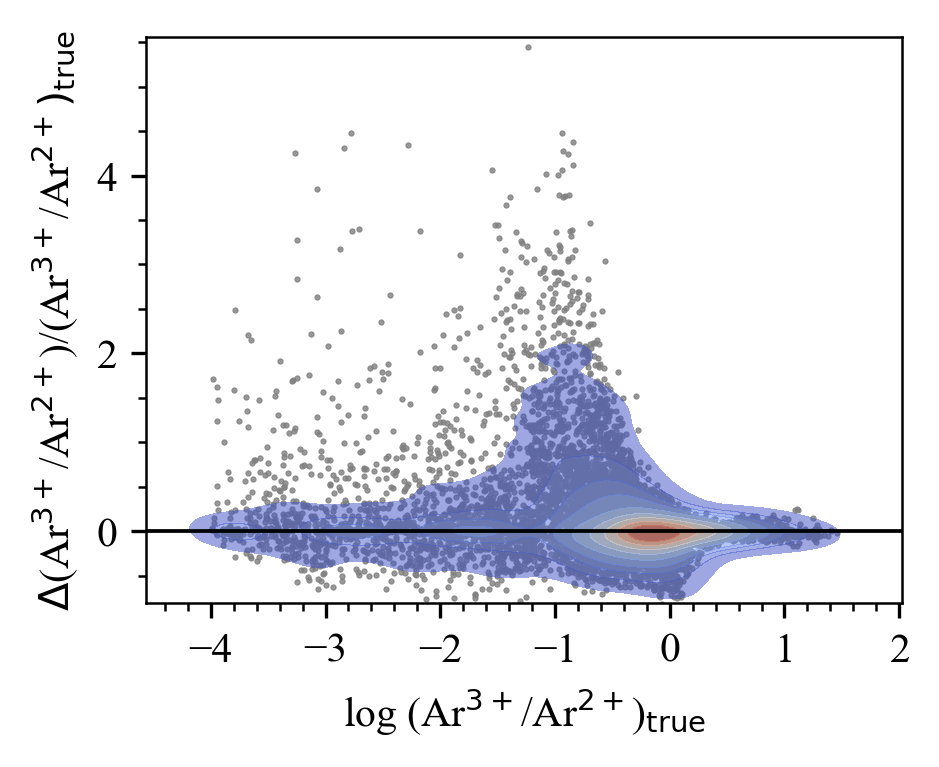}\hspace{0.7cm}
     \includegraphics[width=0.40\textwidth]{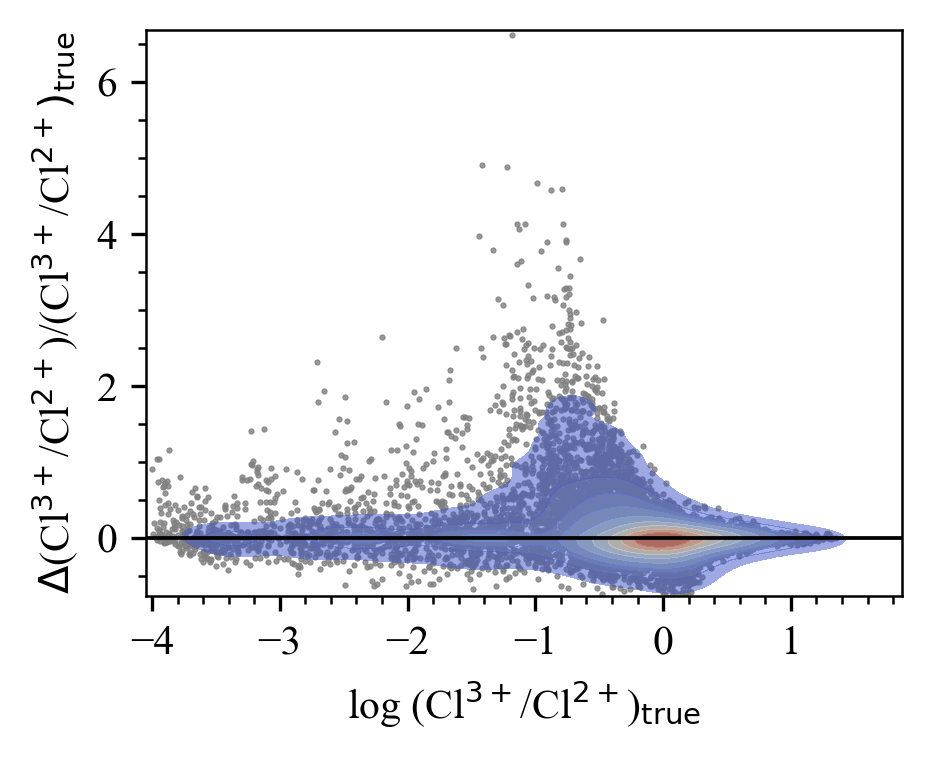}
    \caption{Fractional uncertainty in predicted Ar$^{3+}$/Ar$^{2+}$ (left) and Cl$^{3+}$/Cl$^{2+}$ (right) from the RF regressor as a function of log-scale true ionic fractions of the test set. The plots were constructed in the same manner as Fig.\ref{fig:ML_icfS}.}
    \label{fig:ML_icfS_ionfrac}
\end{figure}

\section{Classifying Planetary Nebulae into basic Young and Old Groups}
\label{appendix:pne_ages}
Based on previous studies on PNe abundances, \citet{peimbert1978chemical} divided PNe into four distinct types, each with unique general properties. Type I PNe are those enriched in He and N abundances \citep[He/H$>0.125$ and $\log$(N/O)$>-0.3$, as defined in][]{peimbert1983type}, and they typically exhibit bipolar morphologies, high central star temperatures, and lower mean scale-heights above the Galactic plane \citep{corradi1995morphological, peimbert1983type}. These properties are consistent with more massive progenitors compared to other PNe nuclei, as younger and more massive stars are likely to form in a more enriched ISM and undergo more efficient ‘dredge-up’ processes \citep{phillips2005mean}, leading to higher He and N abundances. The progenitor masses of Type I PNe are suggested to be greater than $2.4M_{\odot}$ \citep{peimbert1980helium, calvet1983bipolar}, with small variations in the literature, corresponding to an stellar age of about 1~Gyr.

Using AGB evolutionary models that incorporate dust formation in circumstellar envelopes, \citet{ventura2016planetary} investigated the relationship between progenitor masses and chemical composition of the resulting PNe. They found that carbon stars produce C-rich PNe by ejecting their shells, while stars undergoing hot-bottom burning, a phase during which most of their C is converted into N, do not go through the carbon star phase and instead eject a N-rich shell. These findings were used to classify PNe into young and old populations according to their C/H and N/H ratios in \citet{stanghellini2018galactic}, hereafter referred to as SH18. A similar classification scheme using O/H instead of C/H was also derived in SH18 as follows: for the old PNe population with a stellar age greater than 7.5~Gyr, $12+\log$(N/H)$<0.8 \times[12+\log$(O/H)$]+1.4$; and for the young PNe population with the central stars younger than 1~Gyr, $12+\log$(N/H)$>0.6\times[12+\log$(O/H)]$+3.3$.

We consider both PN dating methods as equally reliable. We select a group of young PNe where the two schemes yield non-contradictory results, meaning objects are either classified as young based on the SH18 criteria and are Type I, or are Type~I PNe without an age group determined by the SH18 scheme. Similarly, we choose a group of old PNe that are classified as old using the SH18 criteria and are not Type~I objects. Despite the large uncertainties, which may exceed 3-4 Gyr due to the abundance determination and the dating methods themselves, the young PNe group's progenitor ages are estimated to be less than 1~Gyr, while the old PNe population is expected to descend from those older than 7.5~Gyr. Despite this crude age demarcation, a clear effect in the size and distribution of the sulfur anomaly is shown in Fig.~\ref{fig:SD_age_scatter_hist} in the main body of the text, between the young and old PNe populations.

It is important to note that this age classification relies on the single-star assumption for CSPNe. The presence of companion stars, particularly in close-binary systems, can affect their evolutionary pathways and introduce uncertainty in age estimation \citep[see e.g.][]{soderblom2010ages, stanway2018re}.




\end{document}